\documentclass[journal=jacsat,manuscript=article]{achemso}

\usepackage[version=3]{mhchem} 
\usepackage{adjustbox}
\usepackage{graphics}
\usepackage{adjustbox}
\usepackage[table,xcdraw]{xcolor} 
\usepackage{colortbl}             
\usepackage{booktabs}

\title{CHIPS-FF: Evaluating Universal Machine Learning Force Fields for Material Properties}
\author{Daniel Wines}%
 \email{daniel.wines@nist.gov}
 \affiliation{%
 Material Measurement Laboratory, National Institute of Standards and Technology,
Gaithersburg, MD 20899, USA 
}%

\author{Kamal Choudhary}
 \affiliation{%
 Material Measurement Laboratory, National Institute of Standards and Technology,
Gaithersburg, MD 20899, USA 
}%

\begin{document}

\begin{abstract}
 In this work, we introduce CHIPS-FF (Computational High-Performance Infrastructure for Predictive Simulation-based Force Fields), a universal, open-source benchmarking platform for machine learning force fields (MLFFs). This platform provides robust evaluation beyond conventional metrics such as energy, focusing on complex properties including elastic constants, phonon spectra, defect formation energies, surface energies, and interfacial and amorphous phase properties. Utilizing 16 graph-based MLFF models including ALIGNN-FF, CHGNet, MatGL, MACE, SevenNet, ORB, MatterSim and OMat24, the CHIPS-FF workflow integrates the Atomic Simulation Environment (ASE) with JARVIS-Tools to facilitate automated high-throughput simulations. Our framework is tested on a set of 104 materials, including metals, semiconductors and insulators representative of those used in semiconductor components, with each MLFF evaluated for convergence, accuracy, and computational cost. Additionally, we evaluate the force-prediction accuracy of these models for close to 2 million atomic structures. By offering a streamlined, flexible benchmarking infrastructure, CHIPS-FF aims to guide the development and deployment of MLFFs for real-world semiconductor applications, bridging the gap between quantum mechanical simulations and large-scale device modeling.

\end{abstract}

\textbf{Keywords:} Machine learning force field; deep learning; foundational models; density functional theory; high-throughput; materials discovery; semiconductors

\maketitle

Multiscale modeling in materials science \cite{tadmor2011modeling,xia2022fundamentals} is an essential tool to bridge the gap between atomistic properties, device performance, and manufacturing. Recently, there has been a renewed focus on modeling semiconductor materials, devices, and components. Various approaches to semiconductor modeling can be applied at different scales, including quantum mechanical tools such as density functional theory (DFT)\cite{martin2020electronic}, classical atomistic simulations such as molecular dynamics (MD) \cite{allen2017computer}, technology computer-aided design (TCAD) \cite{vasileska2017computational} and various machine learning (ML) models \cite{ml-mat,Jacobs_2024,marques-ML,choudhary2022recent,han2024aidriveninversedesignmaterials,choudhary2024atomgpt,D3DD00113J}.

In order to better inform semiconductor device models, accurate structure-to-property relationships must be established, which often require more computationally expensive quantum mechanical approaches. Accurate structural relaxations of defect structures and semiconductor interfaces are key to uncovering the underlying physics that govern device performance. For example, defects in semiconductors can introduce energy levels within the band gap, which can impact carrier mobility and concentration and create traps, leading to non-radiative recombination of holes and electrons \cite{gan-defect,BUCKERIDGE2019329,10.1063/5.0135382,10.1063/5.0176333,TURIANSKY2021108056,defect-review}. Semiconductor surfaces and interfaces are especially important to consider since they are the building blocks of devices, impacting charge carrier behavior (band alignment), electrical properties (i.e., Schottky vs. Ohmic), thermal management and device efficiency \cite{10.1063/5.0156437,D4DD00031E,Park_2019,interface-cite,PhysRevB.90.155405,surface-dft}. In addition, the combination of semiconductor materials and high-dielectric amorphous structures are essential for building Metal-Oxide-Semiconductor Field-Effect Transistors (MOSFETs) \cite{HUANG2019616,SIRON2023112192,a-Si,zheng2024abinitioamorphousmaterials,a-bn,a-bn-2}.

Due to the costly scaling of DFT ($N^3$, where $N$ is the number of electrons), high-throughput calculations of larger simulation cells required to study defects, surfaces and interfaces are not feasible. Unfortunately, many of the current classical atomistic interatomic potentials do not have the capability to achieve near-DFT accuracy and capture these complex interactions. In addition, most classical potentials are limited in their transferability to various systems and experimental conditions. Early attempts to create a universal force field for the entire periodic table were limited in their accuracy \cite{rappe-uff}. For this reason, many researchers have explored machine learning force fields (MLFFs) trained on highly-accurate DFT calculations as a viable route to bridge the gap between quantum mechanical accuracy and large-scale atomistic simulations. 

One of the pioneering MLFF architectures was developed by Behler and Parrinello in 2007 using neural networks \cite{behler2007generalized}. Other MLFF methods include Gaussian processes-based Gaussian approximation potentials (GAP) \cite{bartok2010gaussian}, spectral neighbor analysis potential (SNAP) \cite{wood2018extending} and Allegro \cite{musaelian2023learning}. Graph neural network (GNN)-based machine learning force fields (MLFFs) are now considered state-of-the-art due to their superior accuracy and transferability. With the public availability of large DFT databases (such as the Materials Project \cite{10.1063/1.4812323}, JARVIS-DFT \cite{10.1063/5.0159299,jarvis}, OQMD \cite{oqmd-1,oqmd-2}, and Alexandria \cite{https://doi.org/10.1002/adma.202210788,Wang_2023,doi:10.1126/sciadv.abi7948,SCHMIDT2024101560}), several researchers have trained GNN force fields on full DFT databases, encompassing the entire periodic table. These so-called universal/unified/foundational machine learning force fields (uMLFFs) have been successful at reproducing near-DFT accuracy for a wide-range of systems. 

A major advantage of utilizing these pretrained uMLFF architectures is that expensive DFT dataset generation is not an initial requirement. These pretrained uMLFFs can be used in place of DFT or fine tuned on more tailored datasets or tasks. Such uMLFFs have shown remarkable potential for numerous applications including modeling of defects and surfaces \cite{deng2024overcomingsystematicsofteninguniversal,mlff-surface}, disordered alloys \cite{PhysRevMaterials.8.113803}, high-pressure superconductors \cite{wines2024data}, solid-solution energetics and ion migration barriers \cite{deng2024overcomingsystematicsofteninguniversal}, catalysis \cite{wang2024examining}, thermal conductivity \cite{pota2024thermalconductivitypredictionsfoundation}, semiconductor interfaces \cite{D4DD00031E}, guiding inverse materials design \cite{choudhary2024atomgpt,gruver2024fine,cdvae-2dmag} and many other applications in chemistry, physics and materials science \cite{batatia2024foundationmodelatomisticmaterials,neumann2024orbfastscalableneural,fairchem_omat24_2024}. Important quantities such as optimized structure, elastic properties, vibrational properties (phonons), stability, thermal properties, defect formation energy and surface energy can be easily computed from uMLFFs at a significantly lower cost than DFT. Additionally, uMLFF structural relaxations of bulk semiconductors, point defects, surfaces and interfaces can be coupled with other low-cost tools (direct prediction with GNNs \cite{matgl-gap,10.1063/5.0176333,D4DD00031E,10.1063/5.0176333,alignn}, tight binding \cite{PhysRevMaterials.7.044603, tb}) to predict electronic properties. uMLFFs can also be used to perform structure searches (replacing hundreds to tens of thousands of DFT relaxation calculations), which can be followed by a single DFT calculation for electronic properties of the target structure.

One of the first graph-based uMLFF architectures introduced in 2022 was M3GNet from Chen and Ong \cite{m3gnet}. M3GNet has since transformed into MatGL (Materials Graph Library) \cite{matgl_github}, a re-implementation of M3GNet built on the Deep Graph Library (DGL) \cite{wang2020deepgraphlibrarygraphcentric} and PyTorch \cite{paszke2019pytorchimperativestylehighperformance}. M3GNet evolved from the MEGNet (MatErials Graph Network) property prediction model introduced in Ref. \cite{megnet}. M3GNet was trained on $\approx$ 60,000 inorganic materials ($\approx$ 190,000 relaxation steps) in the Materials Project (MP), covering 89 elements of the periodic table \cite{m3gnet}. This dataset is commonly referred to as MPF \cite{m3gnet}. In M3GNet, the angles (representing three-body interactions) are incorporated by aggregating to bonded atoms within the graph convolution steps to update atoms, bonds and properties \cite{m3gnet}. The Atomistic LIne Graph Neural Network (ALIGNN)-FF model \cite{D2DD00096B} was introduced in 2022 and was trained on the JARVIS-DFT dataset of $\approx$ 75,000 materials ($\approx$ 300,000 relaxation steps, ALIGNN-FF DB). ALIGNN-FF evolved from the ALIGNN property prediction model released in 2021 \cite{alignn}. The ALIGNN model achieved significant improvement in property prediction tasks due to the inclusion of bond-angles in terms of line graphs, which previous GNN models lacked \cite{alignn}. The earlier versions of ALIGNN-FF utilized a k-nearest neighbor graph \cite{D2DD00096B}, while the updated version of ALIGNN-FF (2024.12.2) utilizes a radius graph, making it faster and more robust. Another uMLFF that utilizes a crystal graph (atom graph and bond graph) is the Crystal Hamiltonian Graph Neural Network (CHGNet) \cite{chgnet}, which was proposed in 2023. This model was trained on the Materials Project Trajectory (MPtrj) dataset \cite{chgnet}, which includes DFT calculations for $\approx$ 150,000 materials ($\approx$ 1.58 million relaxation steps). CHGNet also incorporates magnetic moments into the training process to enhance the description of chemical reactions \cite{chgnet}.

MACE\cite{Batatia2022mace,Batatia2022Design,batatia2024foundationmodelatomisticmaterials}, an equivariant message passing neural network potential, uses higher order messages (higher than two body) combined with atomic cluster expansion (ACE) \cite{PhysRevB.99.014104,DUSSON2022110946}, which is a method for deriving an efficient body-ordered symmetric polynomial basis to represent functions of atomic neighborhoods. A pre-trained model for the entire periodic table (trained on MPTrj) for MACE was released in late 2023 \cite{batatia2024foundationmodelatomisticmaterials} (referred to as MACE-MP-0). In late 2024, a new pretrained version of MACE (known as MACE-MPA-0) was released, which is trained on a combination of MPTrj and Alexandria \cite{mace_mpa_0}. SevenNet (Scalable EquiVariance Enabled Neural Network) \cite{sevennet}, another equivariant message passing uMLFF, is based on the NequIP \cite{nequip} architecture for building E(3)-equivariant force fields. SevenNet trained on the MPtrj dataset and was released in 2024. In addition to these open source models, there have been proprietary uMLFF models developed including MatterSim (Microsoft) \cite{yang2024mattersimdeeplearningatomistic}, GNoME (Google DeepMind) \cite{gnome}, and PFP (Preferred Networks Inc) \cite{pfp}. A version of MatterSim (MatterSim-v1), which has a similar but slightly modified architecture to M3GNet and is trained on over 17 million structures sampled from Materials Project, Alexandria, and Microsoft's own dataset (consisting of additional structures and MD trajectories at ambient and extreme conditions), was made open source to the community in December 2024 \cite{microsoft_mattersim}. The Orb \cite{orb_models_github} pretrained uMLFF, released in fall 2024, was developed by the startup company Orbital Materials, where superior performance for force and energy predictions were demonstrated at significantly reduced computational cost \cite{neumann2024orbfastscalableneural,orb_technical_blog,matbench_discovery_preprint}. Orb utilizes an attention augmented Graph Network-based Simulator (GNS) \cite{sanchezgonzalez2020learningsimulatecomplexphysics}, which is a type of Message Passing Neural Network (MPNN) \cite{gilmer2017neuralmessagepassingquantum}. In contrast to equivariant models such as MACE and SevenNet, Orb does not utilize equivariant message passing. The open source Orb model (orb-v2) is trained on MPtrj and Alexandria (over 3 million materials and 32 million relaxation steps) \cite{neumann2024orbfastscalableneural}. In late 2024, Meta FAIR Chemistry released the Open Materials 2024 (OMat24) \cite{barrosoluque2024openmaterials2024omat24} models and dataset, which contains over 110 million DFT calculations and pretrained uMLFF models based on the EquiformerV2 \cite{liao2024equiformerv2improvedequivarianttransformer} architecture, which is a highly accurate equivariant transformer model. OMat24 offers various sized models trained on OMat, MPTrj and Alexandria.

Recently, interest in the field of uMLFFs has grown immensely (from academia, government and industry), as seen by the several models above and various use cases of new model architectures. Given the current state of the field, benchmarking the performance of uMLLFs is imperative. Large-scale efforts to benchmark the performance of various pretrained uMLFF architectures have emerged such as Matbench Discovery \cite{riebesell2024matbenchdiscoveryframework,matbench_discovery_preprint} and the JARVIS-Leaderboard \cite{leaderboard,jarvis_leaderboard}, which calculate error metrics and allow for new models to be uploaded as time progresses. Matbench Discovery is an interactive leaderboard platform where uMLFFs are ranked based on accuracy of properties (such as energy and more recently thermal conductivity) and the ability to simulate high-throughput discovery of stable new inorganic materials. These uMLFFs are evaluated on targeted large-scale test datasets (i.e., a test set of 200,000 Materials Project entries) to calculate error metrics \cite{riebesell2024matbenchdiscoveryframework,matbench_discovery_preprint}. The JARVIS-Leaderboard is a flexible benchmarking platform that allows for the evaluation of various machine learning, electronic structure, force-field and quantum computing, and experimental methods. So far, there are over 300 benchmarks and over 2,000 user contributions \cite{leaderboard,jarvis_leaderboard}. A subset of these benchmarks and user contributions are focused on uMLFFs for various test datasets \cite{leaderboard,jarvis_leaderboard}. There have been focused benchmarking efforts for more involved properties beyond energy. Yu et al. \cite{https://doi.org/10.1002/mgea.58} performed a systematic assessment of uMLFFs (M3GNet, ALIGNN-FF, CHGNet and MACE), calculating properties such as optimized structure, formation energy, bulk modulus, and phonon band structure, comparing directly to DFT to calculate error metrics. Benchmarking of more advanced properties such as surface energies, defect formation energies, solid-solution energetics, and ion migration barriers have recently been performed with M3GNet, CHGNet and MACE \cite{deng2024overcomingsystematicsofteninguniversal,mlff-surface}. In addition, a high-level assessment of uMLFF computed thermal conductivity has been conducted with M3GNet, CHGNet, MACE, SevenNet and ORB by Po\'ta et al. \cite{pota2024thermalconductivitypredictionsfoundation} More recently, Zhu et al. utilized M3GNet, CHGNet, MACE, SevenNet, and ORB to accelerate the prediction of CALPHAD-based phase diagrams for complex alloys \cite{zhu2024acceleratingcalphadbasedphasediagram} and Lowe et al. performed in-depth benchmarking of phonons for over 10,000 materials using all of the recent state-of-the-art uMLFFs \cite{loew2024universalmachinelearninginteratomic}. These recent efforts have highlighted the successes and limitations of various uMLFF architectures, and have emphasized the need for a comprehensive publicly available benchmarking platform. Such a platform allows the community to make informed decisions on regarding uMLFFs, outweighing factors such as computational cost versus accuracy. In addition, researchers may find systematic trends in the uMLFF results which can help others to understand the applicability of each uMLFF type. Well-understood systematic trends in DFT functionals (i.e., the underbinding and lattice constant errors of GGA) have been exploited to identify new exfoliable van der Waals (vdW) materials \cite{jarvis-vdw}. In contrast to the Jacob's Ladder metaphor of DFT exchange-correlation functionals \cite{10.1063/1.1390175}, where the ``rungs" of the ladder refer to the increasing complexity (and computational cost) and accuracy of the functionals, there (as of yet) exists no systematic assessment or generalizable statements for uMLFFs. As uMLFFs begin to replace DFT calculations, such analysis is needed to help the community better develop new force fields or fine-tune existing force fields for downstream applications.

In this work, we present CHIPS-FF (Computational High-performance Infrastructure for Predictive Simulation-based Force Fields): a generalized user-friendly open-source benchmarking package to test various uMLFFs for a number of properties beyond energy. We have implemented a streamlined workflow to use uMLFFs for structural relaxation, bulk modulus, elastic properties, point defect formation energy, surface energy, interfacial properties (work of adhesion), molecular dynamics and creation of amorphous structures, all with automated error metrics from JARVIS-DFT. In contrast to large scale benchmarking efforts such as Matbench Discovery \cite{riebesell2024matbenchdiscoveryframework,matbench_discovery_preprint}, which requires users to submit contributions covering large-scale test sets (on the order of 200,000), CHIPS-FF allows for robust benchmarking on smaller datasets and more complex properties. Although our framework is agnostic to the type of material, we chose a set of 104 materials that are most relevant for semiconductor devices as a case study. Additionally, we evaluate force-prediction accuracy of these models for close to 2 million atomic structures from Materials Project and JARVIS-DFT databases. We intend this codebase and benchmarking dataset to benefit the MLFF community and aid in the development of future models.

The CHIPS-FF workflow mainly utilizes the Atomic Simulation Environment (ASE) \cite{ase} and JARVIS-Tools \cite{10.1063/5.0159299,jarvis} to perform atomistic simulations in Python. In terms of pretrained uMLFF models, we utilized ALIGNN-FF: 2.12.2024, CHGNet: 0.3.8, MatGL: 1.1.2 (using both the M3GNet-MP-2021.2.8-PES and M3GNet-MP-2021.2.8-DIRECT-PES \cite{m3gnet-direct} models), MACE: 0.3.10 (using MACE-MP-0 and MACE-MPA-0), SevenNet-0: 0.9.2 (using the 11July2024 model), orb-models: 0.4.1 (orb-v2 and orb-d3-v2, which utilizes the D3 dispersion correction \cite{10.1063/1.3382344}), and MatterSim-v1 (MatterSim-v1.0.0-5M). In addition, we benchmarked a MACE force field trained on the newly released Alexandria dataset presented in Ref. \cite{SCHMIDT2024101560} (this model was trained on over 200,000 2D material calculations with the PBEsol \cite{PhysRevLett.100.136406} functional and is referred to as MACE$^{\textrm{2D}}$ in that work \cite{alexandria_v2_2d_uff}). In our work, we refer to this model as mace-alexandria (which is distinct from mace-mpa). For the OMat24 models from Meta, we tested the three EquiformerV2 (eqV2) models trained on OMat (Small: 31M (million), medium: 86M, and large: 153M in terms of model parameters) and two eqV2 models trained on OMat+MPTrj+Alexandria (Small: 31M and medium: 86M) \cite{fairchem_omat24_2024,barrosoluque2024openmaterials2024omat24}. These models are named as: eqV2\_31M\_omat, eqV2\_86M\_omat, eqV2\_153M\_omat, eqV2\_31M\_omat\_mp\_salex, and eqV2\_86M\_omat\_mp\_salex \cite{fairchem_omat24_2024,barrosoluque2024openmaterials2024omat24}. 

Initial structures were obtained from the JARVIS-DFT database (but not limited by it) and then optimized using the FIRE \cite{PhysRevLett.97.170201} algorithm within ASE. For bulk relaxations, atomic coordinates and lattice vectors were allowed to relax, but for surfaces and vacancy calculations, cell volume was fixed and atomic positions were allowed to relax. The maximum force value (stopping criterion for relaxation) was set to 0.05 eV/\AA. If a calculation did not reach the stopping criterion by 200 steps, it was considered unconverged. If a structure did not reach convergence within 200 steps, the final structure and energy at 200 steps was logged and used for subsequent portions of the workflow. We tested both the \texttt{FrechetCellFilter} (main text) and \texttt{ExpCellFilter} (in SI) within ASE  \cite{ASEFilters}. The bulk modulus was obtained by applying isotropic strain to the system from -6 $\%$ to +6 $\%$ and using a Murnaghan fit for the equation of state. The elastic tensor was computed using the Elastic package \cite{jochym_elastic,elastic1,elastic2}. For bulk modulus and elastic tensor calculations, the conventional cell was used. We utilized the phonopy \cite{phono3py,phonopy-phono3py-JPCM} package to perform simulations of the phonon spectrum using the finite displacement method \cite{PhysRevB.84.094302}, where a 2 x 2 x 2 supercell was constructed from the initial unit cell. Displacement values of 0.2 \AA, 0.05 \AA, 0.01 \AA, and 0.001 \AA\space were benchmarked. From these phonopy results, we also extracted the Helmholtz free energy (and zero point energy at 0 K), entropy and heat capacity as a function of temperature. We have added capabilities to CHIPS-FF to compute the thermal expansion of each material at constant pressure and the lattice thermal conductivity, utilizing the quasi-harmonic approximation \cite{PhysRevB.81.174301} implemented in phonopy and phono3py \cite{phono3py,phonopy-phono3py-JPCM} respectively (these simulations are ongoing and will be discussed in future work). JARVIS-Tools was used to automatically generate supercells for neutral point defects (vacancies) and non-polar surfaces (both of which were later relaxed with uMLFFs). In this work, we obtained initial defect structures from the relaxed structures in the JARVIS-DFT vacancy database from Ref. \cite{10.1063/5.0135382}. This allowed us to make a 1:1 comparison between vacancy formation energies (calculated with DFT and uMLFFs) for the exact same crystal structure and supercell size (avoiding finite-size scaling effects). In Ref. \cite{10.1063/5.0135382}, these defect supercells were constructed from the conventional cell and enforced to have a lattice vector c of at least 8 \AA\space, where single vacancies were created based on Wycoff-position information for each atomic species in the crystal. The following formula is used to obtain the vacancy formation energy ($E_{\text{vac}}$):  

\begin{equation}
E_{\text{vac}} = E_{\text{defect}} - E_{\text{bulk}} + \mu
\end{equation}
where $E_{\text{defect}}$ is the total energy of the defect supercell, $E_{\text{bulk}}$ is the energy of the bulk structure (no defect) and $\mu$ is the chemical potential of the missing atom. The lowest energy crystal structure of the elemental solid (from JARVIS-DFT) is used to calculate the chemical potential with each respective uMLFF (where the structure is fully re-relaxed). Surfaces were generated for miller indices of [1, 0, 0], [1, 1, 1], [1, 1, 0], [0, 1, 1], [0, 0, 1], and [0, 1, 0], skipping any polar surfaces in the surface relaxation. Each surface contained at least 4 layers and 18 \AA\space of vacuum. The surface energy of each non-polar surface was calculated as:
\begin{equation}
\gamma = \frac{E_{\text{surface}} - N \cdot E_{\text{bulk}}}{2A}
\end{equation}
where $E_{\text{surface}}$ is the energy of the surface structure, $E_{\text{bulk}}$ is the energy of the bulk structure, $N$ is the number of bulk unit cells in the surface, and $A$ is the cross sectional area of the surface. In order to study amorphous materials, we have implemented finite-temperature molecular dynamics into our workflow to perform melt/quench simulations using Berendsen NVT (Number of particles, Volume, Temperature ensemble) dynamics within ASE. For our case study of amorphous Si, we used a timestep of 1 femtosecond and ran the simulation at 3500 K for 10 ps (melt) and 300 K for 20 ps (quench). For benchmarking calculations to compare uMLFF results to, we performed similar ab initio MD (AIMD) calculations with the Vienna Ab initio Simulation Package (VASP) code, using projector augmented wave (PAW) pseudopotentials \cite{PhysRevB.54.11169,PhysRevB.59.1758} and the vdW-DF-optB88 functional.  These gamma-point AIMD calculation used a plane wave cutoff energy of 500 eV and the Nos\'e-Hoover\cite{10.1063/1.447334,10.1143/PTPS.103.1,PhysRevA.31.1695} thermostat to run NVT simulations at 2000 K for 5 ps (melting), and 300 K for 5 ps (at a timestep of 1 femtosecond). In order to perform calculations for material interfaces, we utilized InterMat \cite{D4DD00031E}, a newly developed package for the generation and calculation of interface structures (substrate plus film). The initial interface is obtained by creating a superlattice or alternating slab junction (ASJ) structure (without vacuum padding) \cite{PhysRevB.35.8154,offset} and then using the Zur algorithm \cite{10.1063/1.333084} to obtain the best candidate interface. From here, we computed the relative alignment between the film and substrate in the in-plane (xy) direction by performing a grid search with a 0.05 fractional spacing interval with each respective uMLFF model. In addition to determining the optimal in-plane orientation between the substrate and film, this gives us a qualitative estimate of how smooth the potential energy surface is at the interface for each pretrained uMLFF. From this calculation, we also determined the work of adhesion ($W_{\text{ad}}$) at the interface with: 
\begin{equation}
W_{\text{ad}} = \gamma_{\text{film}} + \gamma_{\text{substrate}} - \gamma_{\text{interface}}
\end{equation}
where $\gamma_{\text{film}}$ is the surface energy of the film, $\gamma_{\text{substrate}}$ is the surface energy of the substrate and $\gamma_{\text{interface}}$ is the interfacial energy calculated as:
\begin{equation}
\gamma_{\text{interface}} = \frac{E_{\text{interface}} - E_{\text{film}} - E_{\text{substrate}}}{A}
\end{equation}
where $E_{\text{interface}}$ is the total energy of the interface, $E_{\text{film}}$ is the total energy of the film, $E_{\text{substrate}}$ is the total energy of the substrate and $A$ is the cross sectional area. 

\begin{figure}[h]
    \centering
    \includegraphics[trim={0. 0cm 0 0cm},clip,width=1.0\textwidth]{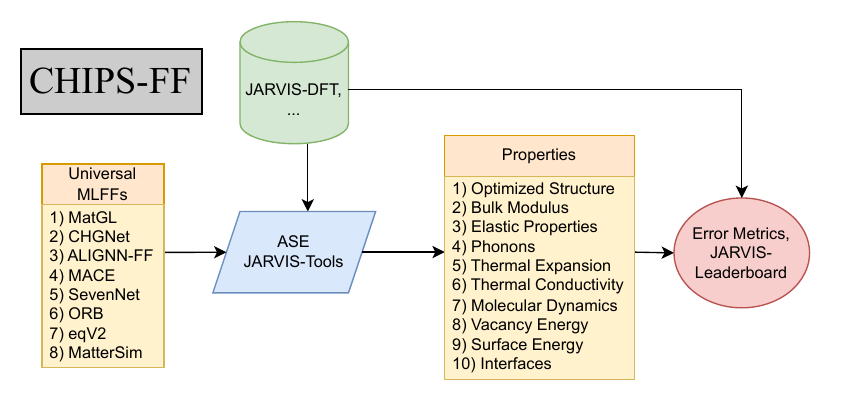}
    \caption{A full schematic of the CHIPS-FF workflow. }
    \label{workflow}
\end{figure}

A majority of the reference data used to compare our uMLFF results to were obtained from the JARVIS-DFT database, which contains relaxed structural data (primitive lattice vectors, volume), elastic properties (bulk modulus, elastic tensor) and phonon band structure \cite{10.1063/5.0159299,jarvis}. Within JARVIS-DFT, there exists a database of single vacancy calculations \cite{10.1063/5.0135382} and surfaces \cite{D4DD00031E} that were both used for benchmarking these uMLFFs. All of the DFT data in JARVIS was computed with the vdW-DF-optB88 \cite{Klimes_2010} functional, which is a modified Generalized Gradient Aprroximation (GGA) functional designed to explicitly include van der Waals (vdW) interactions. This is slightly different than other materials repositories such as the Materials Project (MPtrj), which mostly contains DFT calculations performed with the Perdew Burke Ernzerhof (PBE) \cite{PhysRevLett.77.3865} GGA functional. 
Despite mostly comparing to JARVIS-DFT results (vdW-DF-optB88), the CHIPS-FF workflow has a flexible framework to use any DFT dataset as a ground truth (i.e., the Materials Project, Alexandria, a users own DFT dataset, etc.). In addition, the platform is flexible enough to compare uMLFF results directly to available experimental values. Of course, comparing the results of a quantity computed with a uMLFF to a DFT result with an arbitrary exchange-correlation functional and varying convergence criteria may result in biased or inconclusive error metrics. In this case, it may be suitable to compare uMLFF results directly to available experimental data, especially if the uMLFF prediction is closer to ``reality" than the DFT calculation it is being evaluated against. Although it is possible to assess the accuracy of MLFFs by comparing results directly to DFT or experiment, there is a lack of explicit uncertainty quantification (UQ) for most MLFFs, including the universal pretrained MLFFs mentioned in this work. A robust UQ method should account for uncertainty arising from measurement noise (aleatoric) and uncertainty in predictions arising from model error (epistemic). Due to the fact that the DFT training data does not possess aleatoric error, most of the uncertainty for MLFFs is epistemic, which can be due to scarcity of data, limitations of model architecture, or poor parameter optimization within the model. Recently, different epistemic UQ methods such as ensemble-based uncertainty \cite{blundell2017simple}, deep evidential regression \cite{amini2020deep}, mean-variance estimation \cite{nix1994estimating}, Gaussian mixture models \cite{zhu2023fast} have been implemented and benchmarked for atomistic MLFFs \cite{uq,vita2024ltaufflosstrajectoryanalysis}. It is evident that low-cost and widespread implementations of UQ for universal pretrained uMLFFs is a necessary future development.

\begin{table}[htbp]
  \centering
  \caption{The percentage of unconverged structural relaxations for bulk, surface, and vacancy calculations for each uMLFF (relaxed using the \texttt{FrechetCellFilter}). }
  \label{tab:bulk_surface_vacancy}
  \begin{adjustbox}{max width=\textwidth}
    \begin{tabular}{lccc}
      \toprule
      \textbf{uMLFF Type} & \textbf{Bulk} & \textbf{Surface} & \textbf{Vacancy} \\
      \midrule
      alignn\_ff & 6 & 44 & 35 \\
      chgnet & \cellcolor[HTML]{92D050}0 & \cellcolor[HTML]{92D050}0 & \cellcolor[HTML]{92D050}0\\
      eqV2\_153M\_omat & 1 & 2 & \cellcolor[HTML]{92D050}0\\
      eqV2\_31M\_omat & \cellcolor[HTML]{92D050}0 & \cellcolor[HTML]{92D050}0 & \cellcolor[HTML]{92D050}0\\
      eqV2\_31M\_omat\_mp\_salex & \cellcolor[HTML]{92D050}0 & \cellcolor[HTML]{92D050}0 & \cellcolor[HTML]{92D050}0\\
      eqV2\_86M\_omat & \cellcolor[HTML]{92D050}0 & 1 & \cellcolor[HTML]{92D050}0\\
      eqV2\_86M\_omat\_mp\_salex & \cellcolor[HTML]{92D050}0 & \cellcolor[HTML]{92D050}0 & \cellcolor[HTML]{92D050}0 \\
      mace & \cellcolor[HTML]{92D050}0 & 1 & \cellcolor[HTML]{92D050}0\\
      mace-alexandria & \cellcolor[HTML]{92D050}0 & 1 & \cellcolor[HTML]{92D050}0\\
      mace-mpa & \cellcolor[HTML]{92D050}0 & \cellcolor[HTML]{92D050}0 & \cellcolor[HTML]{92D050}0\\
      matgl & \cellcolor[HTML]{92D050}0 & \cellcolor[HTML]{92D050}0 & \cellcolor[HTML]{92D050}0\\
      matgl-direct & \cellcolor[HTML]{92D050}0 & \cellcolor[HTML]{92D050}0 & \cellcolor[HTML]{92D050}0 \\
      mattersim & \cellcolor[HTML]{92D050}0 & \cellcolor[HTML]{92D050}0 & \cellcolor[HTML]{92D050}0\\
      orb-d3-v2 & \cellcolor[HTML]{92D050}0 & \cellcolor[HTML]{92D050}0 & \cellcolor[HTML]{92D050}0\\
      orb-v2 & \cellcolor[HTML]{92D050}0 & \cellcolor[HTML]{92D050}0 & \cellcolor[HTML]{92D050}0\\
      sevennet & \cellcolor[HTML]{92D050}0 & \cellcolor[HTML]{92D050}0 & \cellcolor[HTML]{92D050}0\\
      \bottomrule
    \end{tabular}
  \end{adjustbox}
\end{table}

\begin{figure}[h]
    \centering
    \includegraphics[trim={0. 0cm 0 0cm},clip,width=1.0\textwidth]{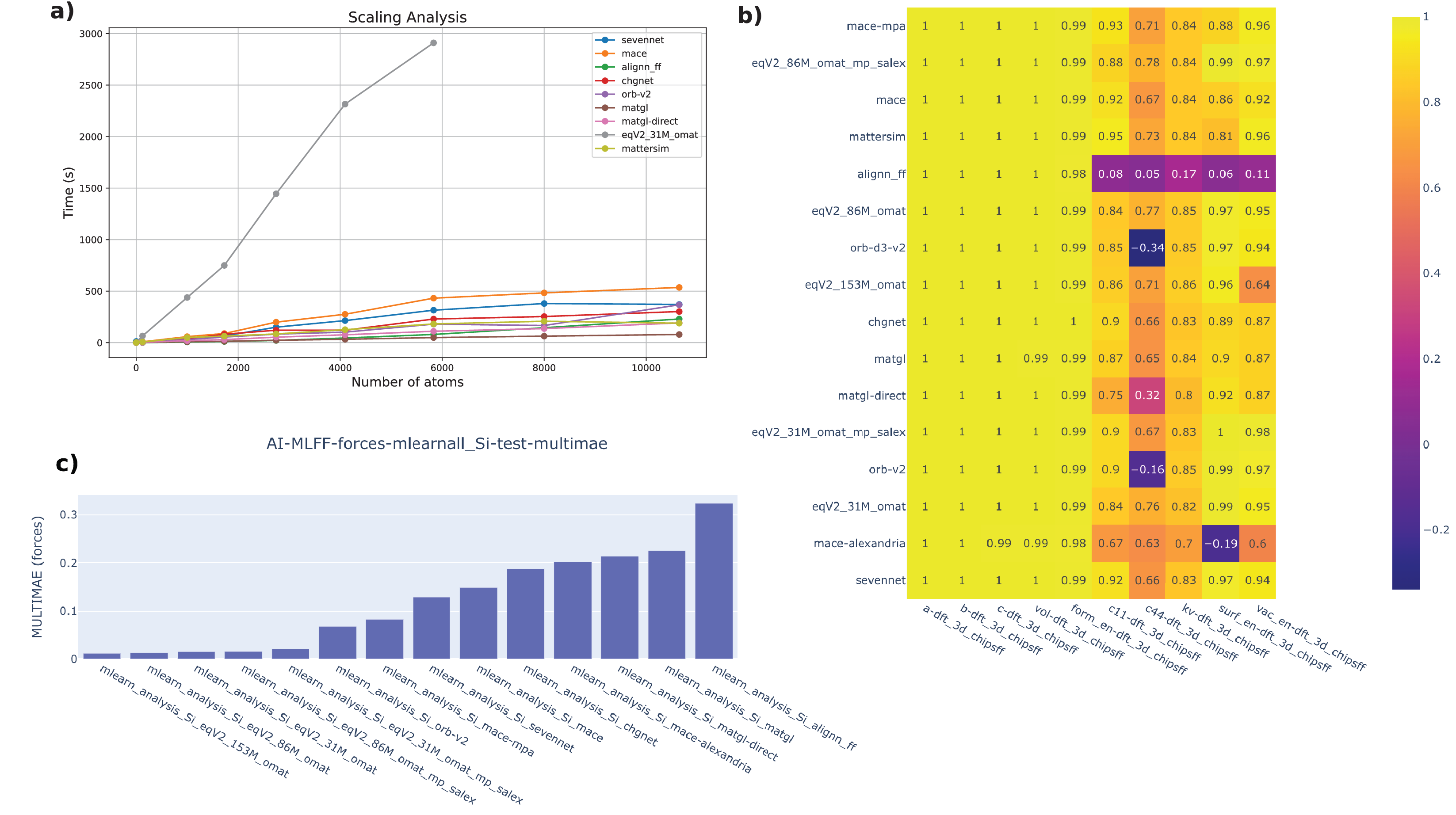}
    \caption{a) Scaling analysis of various uMLFF up to 10,000 atoms (for a supercell of Cu), b) an example JARVIS-Leaderboard entry for the force MAE of the MLEARN dataset, c) an example of the interactive error metrics within the JARVIS-Leaderboard (Pearson correlation coefficient).  }
    \label{scalingandlb}
\end{figure}

\begin{table}[htbp]
  \centering
  \caption{The mean absolute error (MAE) and total computational time for a variety of material properties calculated with each uMLFF type (relaxed using the \texttt{FrechetCellFilter}). The MAE is calculated with respect to JARVIS-DFT data. Properties include (in order) lattice constants a, b, c, volume, bulk modulus, and C11 and C44 components of the elastic tensor. The computational time is measured per CPU. An interactive and up-to-date version of the table is available in the JARVIS-Leaderboard. The green indicates the best performing model.}
  \label{tab:error_metrics}
  \begin{adjustbox}{max width=\textwidth}
    \begin{tabular}{lcccccccccc}
      \toprule
      \textbf{uMLFF Type} & \textbf{err\_a} & \textbf{err\_b} & \textbf{err\_c} & \textbf{err\_vol} & \textbf{err\_kv} & \textbf{err\_c11} & \textbf{err\_c44}  & \textbf{Time} \\
             & \AA & \AA & \AA & \AA$^{3}$ & GPa & GPa & GPa  & seconds \\
      \midrule
      
      alignn\_ff                & \cellcolor[HTML]{92D050}0.011  & \cellcolor[HTML]{92D050}0.011 & \cellcolor[HTML]{92D050}0.014 & \cellcolor[HTML]{92D050}0.42 & 135 & 196 & 74 & 33,081.6  \\
      chgnet                     &  0.036 & 0.038 & 0.072 & 3.29 & 89 & 59 & 46 &  14,102.6  \\
      eqV2\_153M\_omat           & \cellcolor[HTML]{92D050}0.022  & \cellcolor[HTML]{92D050}0.025 & 0.055 & 2.54 & 117 & 75 & \cellcolor[HTML]{92D050}32 &  304,173.9  \\
      eqV2\_31M\_omat            &  \cellcolor[HTML]{92D050}0.015 & \cellcolor[HTML]{92D050}0.016 & 0.052 & 2.25 & 114 & 78 & 37 & 81,420.3  \\
      eqV2\_31M\_omat\_mp\_salex & \cellcolor[HTML]{92D050}0.016  & \cellcolor[HTML]{92D050}0.017 & 0.043 & \cellcolor[HTML]{92D050}1.73 & 112 & 46 & 41 &  65,971.6 \\
      eqV2\_86M\_omat            &  \cellcolor[HTML]{92D050}0.022 & \cellcolor[HTML]{92D050}0.025 & 0.047 & 2.26 & 118 & 86 & 35 &  218,168.4 \\
      eqV2\_86M\_omat\_mp\_salex &  \cellcolor[HTML]{92D050}0.016 & \cellcolor[HTML]{92D050}0.017 & 0.054 & 2.23 & 113 & 43 & 35 &  146,318.1  \\
      mace                       & 0.027  & 0.027 & 0.075 & 2.85 & 94 & 42 & 38 & 27,620.4 \\
      mace-alexandria            & 0.068  & 0.072 & 0.191 & 5.65 & \cellcolor[HTML]{92D050}78 & 69  & 45 &  62,932.5 \\
      mace-mpa                   &  0.040 & 0.042 & 0.084 & 3.58 & 110 & \cellcolor[HTML]{92D050}35 & \cellcolor[HTML]{92D050}34 &  27,255.2  \\
      matgl                      & 0.043  & 0.046 & 0.106 & 4.23 & \cellcolor[HTML]{92D050}83 & 64 & 41 & 10,446.1  \\
      matgl-direct               & 0.043  & 0.042 & 0.089 & 3.00 & \cellcolor[HTML]{92D050}84 & 72 & 50 & 13,420.6   \\
      mattersim                  & 0.027  & 0.028 & 0.089 & 3.11 & 110 & \cellcolor[HTML]{92D050}32 & \cellcolor[HTML]{92D050}32 & 19,047.3 \\
      orb-d3-v2                  & \cellcolor[HTML]{92D050}0.025  & \cellcolor[HTML]{92D050}0.025 & \cellcolor[HTML]{92D050}0.032 & \cellcolor[HTML]{92D050}1.60 & 130 & 77 & 74 & 15,118.9 \\
      orb-v2                     & \cellcolor[HTML]{92D050}0.015  & \cellcolor[HTML]{92D050}0.016 & 0.044 & \cellcolor[HTML]{92D050}1.91 &  118 & 81 & 74 & 12,184.7   \\
      sevennet                   &  0.028 & 0.030 & 0.094 & 3.39 & 103 & 42 & 37 &  17,739.3  \\
      \bottomrule
    \end{tabular}
  \end{adjustbox}
\end{table}

\begin{figure}[htbp]
    \centering
    \includegraphics[trim={0. 0cm 0 0cm},clip,width=1.0\textwidth]{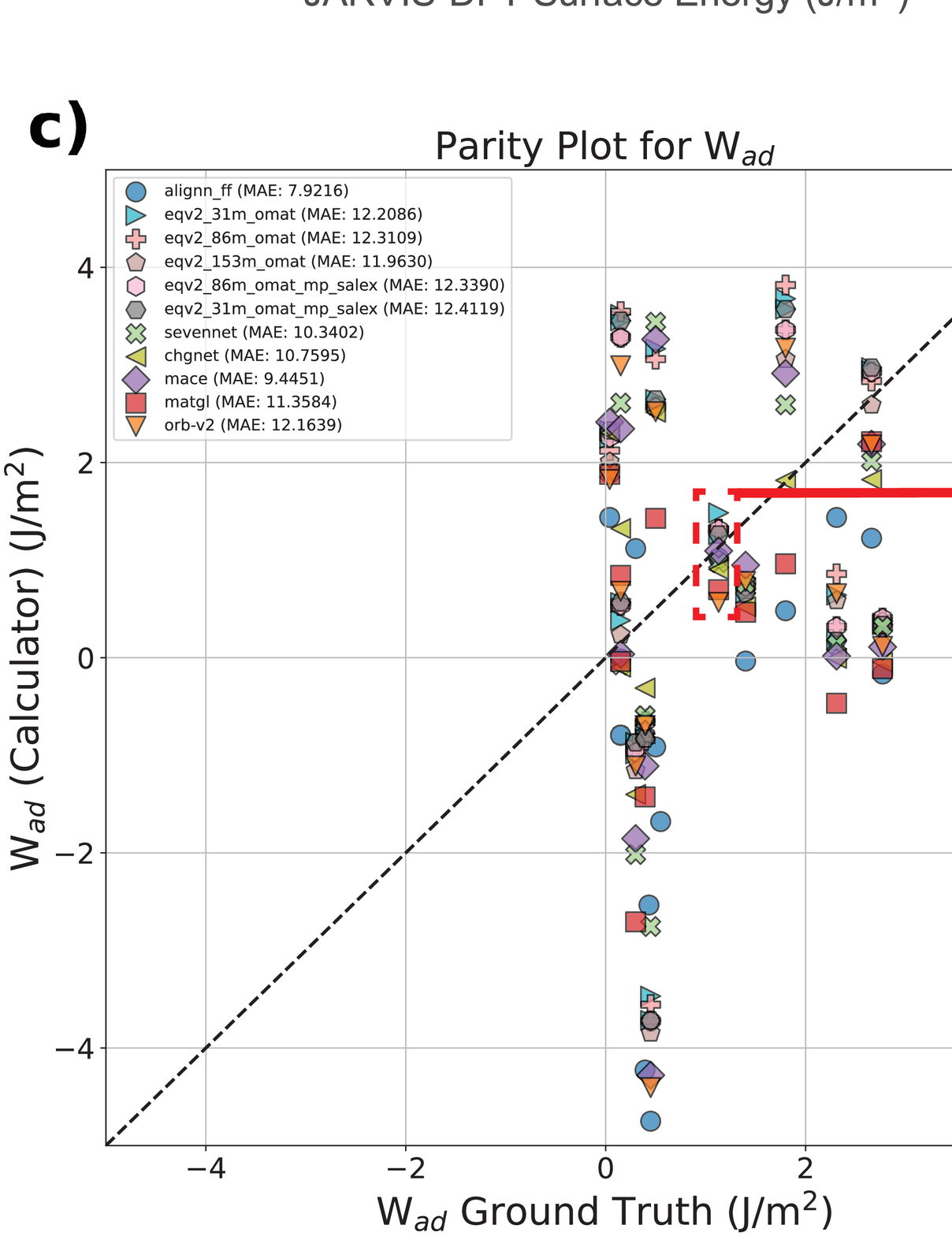}
    \caption{Parity plots for a) surface energy (in J/m$^2$) and b) vacancy formation energy (in eV) for each uMLFF type with respect to JARVIS-DFT data (MAE is given in the inset and the best performing models are included in the plots). c) Depicts the parity plot for work of adhesion (W$_{ad}$ in J/m$^2$) for several material interfaces and MAE with respect to previous calculated and experimental data (taken from Ref. \cite{intermat_interface_2024}). The parity plot is zoomed in to focus on a subset of calculations, but the MAE is computed for the entire test set. To the right of the parity plot, we focus on the best performing W$_{ad}$ prediction for the Al(111)Al$_2$O$_3$(001) interface, showing the xy grid search and optimal in-plane interface configurations with each uMLFF.}
    \label{mlff-parity}
\end{figure}

\begin{figure}[htbp]
    \centering
    \includegraphics[trim={0. 0cm 0 0cm},clip,width=1.0\textwidth]{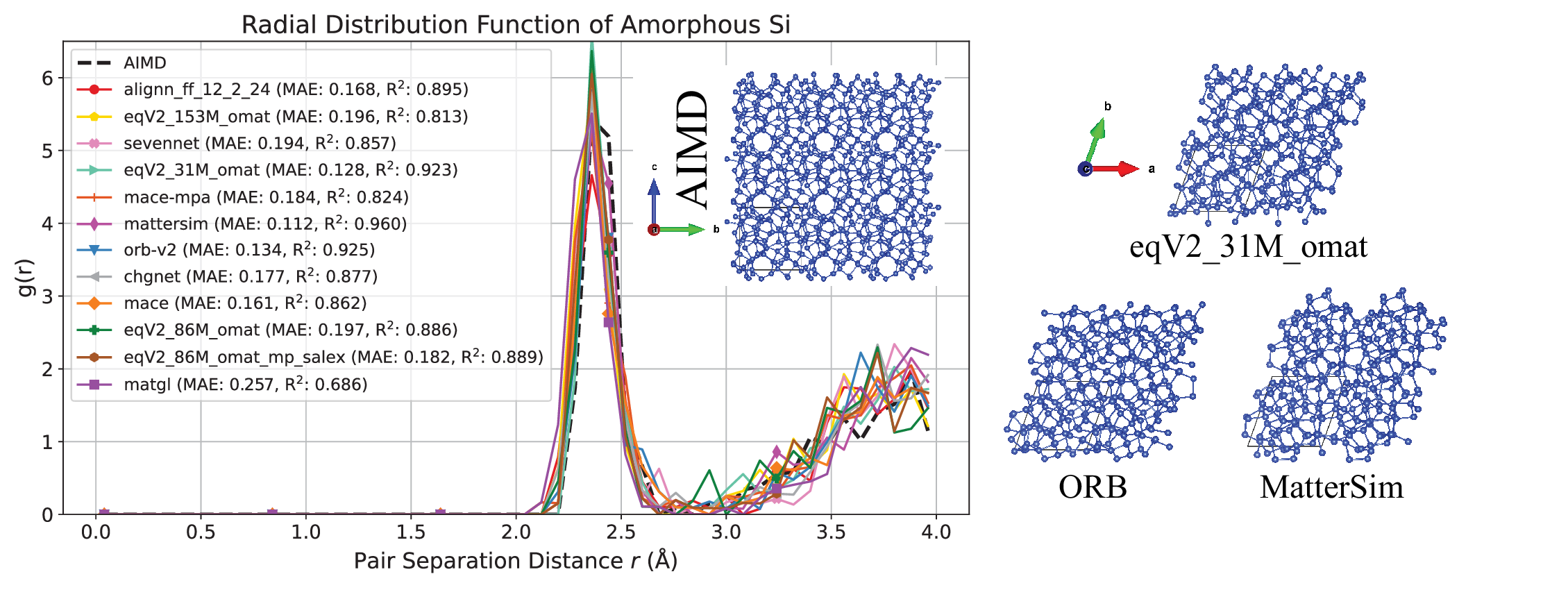}
    \caption{The calculated radial distribution function ($g(r)$) as a function of pair separation distance computed with several uMLFF models for amorphous Si (by performing melt/quench simulations). Ab initio (AIMD) results are given as a ground truth benchmark (black dotted line) and the MAE and R$^2$ with respect to AIMD is reported in the inset. The atomic structure for amorphous Si is given for AIMD and the best performing uMLFF models (MatterSim, eqV2\_31M\_omat and ORB).  }
    \label{mlff-md}
\end{figure}

Fig. \ref{workflow} depicts a full schematic of the CHIPS-FF workflow. In this workflow, initial structures are taken from the JARVIS-DFT database. JARVIS-Tools is used to pull structures from the JARVIS-DFT database and generate supercell surface and defect structures prior to atomistic simulations. Various uMLFF calculators are accessed through ASE. So far, we have implemented MatGL, CHGNet, ALIGNN-FF, MACE, SevenNet, ORB, MatterSim, and eqV2 (OMat24). As new pretrained force fields are developed over time, they will be added to the workflow. In addition, CHIPS-FF offers the flexibility for the user to add their own force field or property to the pipeline. Once each uMLFF is loaded and run through the ASE calculator, a variety of properties can be computed including: optimized structure, bulk modulus (equation of state fitting), elastic tensor (elastic package), phonons (phonopy), thermal expansion (phonopy), thermal conductivity (phono3py), molecular dynamics, vacancy formation energy, surface energy, and interface properties (InterMat). As these calculations are carried out, detailed logging information is saved, including the computational timing for each stage of the calculation and whether or not each stage converged. Once these material properties are computed with each uMLFF, they are cross checked with entries in the JARVIS-DFT database and the mean absolute error (MAE) for each property is computed. For spectral quantities such as the phonon band structure, we computed the MAE for each band along the q-point path using:
\begin{equation}
\text{MAE} = \frac{1}{N_q} \sum_{q\nu} \left| \omega_{q\nu}^{\text{uMLFF}} - \omega_{q\nu}^{\text{DFT}} \right|
\end{equation}
where $\omega_{q\nu}$ is the phonon energy (in cm$^{-1}$) at each q-point and each branch index ($\nu$) (calculated with uMLFF and DFT from JARVIS-DFT) and $N_q$ represents the number of wavevectors along the q-path. In addition to automatically calculating the error metrics, our workflow will create entries for the JARVIS-Leaderboard in the appropriate format (.json.zip and .csv.zip files with the proper naming convention). From here, these entries can be directly uploaded to the leaderboard and compared to other benchmarks. Although we use JARVIS-DFT for the initial structures and calculation of the error metrics, our CHIPS-FF platform is flexible to use other datasets for initial structures and ground truth comparison. Our CHIPS-FF package offers easy-to-use command line tools, which can easily be parallelized to simultaneously run hundreds or even thousands of materials/uMLFFs on CPU and GPU.

For our benchmarking study, we chose a set of 104 materials most commonly found in semiconductor devices and interfaces. This test set contains metals, semiconductors and insulators to be representative of the various parts and interfaces of integrated circuits (semiconductor/insulator interfaces, semiconductor/metal contacts, etc.). Fig. S1 depicts various distributions of properties of the test set, including space group, band gap (calculated with vdW-DF-optB88), atomic number, chemical formula type, crystal system, and Wyckoff site, and dimensionality, all obtained from JARVIS-DFT. In addition, Table S1 and S2 give detailed information of the chemical composition, crystal structure and band gap for each material, where we see that the set contains 28 metals (zero band gap), 14 insulators (band gap above 3 eV) and 62 semiconductors (band gap between 0 eV and 3 eV). In addition, we see from Fig. S1 that our test set represents a diverse set of materials ideal for testing our uMLFF workflow.

When considering the use of a pretrained uMLFF, the most important considerations should be accuracy, rate of convergence, scalability, and computational cost. After running the full CHIPS-FF workflow for the test set of 104 materials, we performed an analysis of which structural relaxation calculations reached convergence (force-maximum threshold of 0.05 eV/\AA\space within 200 steps) for bulk materials, surfaces and vacancies. This analysis is presented in Table \ref{tab:bulk_surface_vacancy} (\texttt{FrechetCellFilter}) and Table S3 (\texttt{ExpCellFilter}), where the percentage of unconverged results is depicted for each calculation type. Aside from ALIGNN-FF, all other uMLFF have a near-perfect convergence rate for bulk structures. With regards to surfaces and defects, the convergence rate is more variable when using the \texttt{ExpCellFilter}, but near-perfect when using the \texttt{FrechetCellFilter} (with the exception of ALIGNN-FF). Interestingly, the OMat24 models significantly vary in terms of convergence for surfaces depending on training data (for the \texttt{ExpCellFilter}). Specifically, the OMat24 models trained on OMat+MPTrj+sAlexandria have a significantly higher rate of convergence when compared to the models solely trained on OMat (see Table S3). These results (Table \ref{tab:bulk_surface_vacancy} vs. Table S3) demonstrate the robustness of the \texttt{FrechetCellFilter} in ASE for these uMLFF models.

For each of the 104 structures, we performed structural relaxations, fitting of the energy vs. volume curve (equation of state), calculation of the elastic tensor, calculation of the phonon band structure, relaxation of all nonpolar surfaces (among [1, 0, 0], [1, 1, 1], [1, 1, 0], [0, 1, 1], [0, 0, 1], and [0, 1, 0]) , and relaxation of each type of point defect in the material. The computational timing and errors in the lattice constants, and elastic properties for each uMLFF model are depicted in Table \ref{tab:error_metrics} (computed with the \texttt{FrechetCellFilter}). An interactive version of these results is available on the JARVIS-Leaderboard platform (see Fig. \ref{scalingandlb}b) for Pearson Correlation Coefficient metric), where it will be continuously updated as new uMLFF models are developed and released. In addition, structural results computed using the \texttt{ExpCellFilter} are depicted in Table S4. Unsurprisingly, we observe more accurate lattice parameters computed with the \texttt{FrechetCellFilter} due to its robustness in optimizing structural parameters (see Table \ref{tab:error_metrics} vs. Table S4). In Table \ref{tab:error_metrics}, we see that all uMLFF models are able to compute reasonably accurate structural parameters. ALIGNN-FF does an excellent job of simultaneously capturing a, b, and c. This is expected due to the fact that ALIGNN-FF was trained on the JARVIS-DFT dataset (vdW-DF-optB88) and the target/``ground truth", in addition to the initial structures, are from JARVIS-DFT. The OMat24 models perform exceedingly well, but are the most computationally expensive. The ORB models also perform exceedingly well for structural relaxation and have the advantage of being close to an order of magnitude more computationally efficient. One interesting trend we observe in our results for this set of 104 materials is that the error for lattice constant c is substantially higher than a and b. This can be due in part to the test set containing a significant portion ($\approx$ 10 $\%$) of vdW bonded (``2D-bulk-like") layered structures (see Fig. S1). Since most of these uMLFF were trained on PBE data without vdW corrections (with the exception of ALIGNN-FF), we can expect larger errors for vdW systems. For example, we observe significant errors for hexagonal BN (JVASP-62940) when relaxed with orb-v2, where we observe over 10 $\%$ error in c. These errors are drastically more significant when using the \texttt{ExpCellFilter} along with orb-v2, where we observed a 130 $\%$ error in c for JVASP-62940 (see Table S4 for total MAE). Fortunately, some of these models such as MACE and ORB have explicit dispersion corrections that can be added to more accurately address vdW interactions \cite{Batatia2022mace,Batatia2022Design,batatia2024foundationmodelatomisticmaterials,neumann2024orbfastscalableneural}. To test this, we ran the workflow for dispersion-corrected orb-d3-v2. For orb-d3-v2, we observe a reduction in error for c and volume (see Table \ref{tab:error_metrics}). For JVASP-62940 relaxed with orb-d3-v2, the percent error in c is reduced to 0.02 $\%$. This emphasizes the importance of dispersion corrections in uMLFF architectures for vdW bonded materials.

Fitting the equation of state to obtain bulk modulus and calculating the elastic tensor prove to be a more challenging task for these uMLFF models, emphasizing difficulties in modeling the potential energy landscape far from equilibrium. We found that models such as MACE-Alexandria and MatGL have superior performance in predicting the energy vs. volume curves compared to the OMat and ORB models. For contributions to the elastic tensor (C11 and C44), we also observe relatively high errors. We see that only a few models can simultaneously predict C11 and C44 with reasonable accuracy (MAE less than 50 GPa). Interestingly, we found that eqV2\_31M\_omat, eqV2\_86M\_omat and eqV2\_153M\_omat can predict C44 with reasonable accuracy, but have much higher prediction errors for C11, while eqV2\_31M\_omat\_mp\_salex and eqV2\_86M\_omat\_mp\_salex do a better job at predicting C11 and C44 simultaneously. We observe that MACE-MPA-0 and MatterSim give the best results for simultaneous C11 and C44 predictions.

\begin{table}[h]
    \centering
    \begin{tabular}{lcccc}
        \toprule
        \textbf{Method} & \multicolumn{4}{c}{\textbf{Displacements (\AA)}} \\
        \cmidrule(lr){2-5}
        & \textbf{0.2} & \textbf{0.05} & \textbf{0.01} & \textbf{0.001} \\
        \midrule
        alignn\_ff & 155 & 155 & 157 & 164 \\
        chgnet & 70 & 75 & 83 & 91 \\
        eqV2\_153M\_omat & 49 & 47 & 49 & 75 \\
        eqV2\_31M\_omat & 48 & 46 & 53 & 73 \\
        eqV2\_31M\_omat\_mp\_salex & 48 & 46 & 112 & 219 \\
        eqV2\_86M\_omat & 51 & 47 & 49 & 77 \\
        eqV2\_86M\_omat\_mp\_salex & 48 & 46 & 103 & 208 \\
        mace & 58 & 60 & 60 & 60 \\
        mace-alexandria & 82 & 84 & 85 & 85 \\
        mace-mpa & 50 & 50 & 50 & 50 \\
        matgl & 82 & 93 & 94 & 94 \\
        matgl-direct & 86 & 101 & 102 & 101 \\
        mattersim & 48 & 47 & 47 & 47 \\
        orb-d3-v2 & 53 & 50 & 85 & 181 \\
        orb-v2 & 49 & 50 & 108 & 201 \\
        sevennet & 54 & 55 & 56 & 56 \\
        \bottomrule
    \end{tabular}
    \caption{MAE for phonon band structure (in cm$^{-1}$) computed with different values of displacement.}
    \label{tab:displacements}
\end{table}

Out of the 104 materials in our test set, there were 84 matching entries for DFT phonon calculations in JARVIS, from which the composite MAE of the phonon band structure was computed in Table \ref{tab:displacements}. We performed these phonon calculations with the uMLFFs at various displacements in the finite-displacement calculation of force constants. We found that models such as ALIGNN-FF and MatGL have relatively large errors in phonon predictions. For certain uMLFFs, we observe a consistent phonon MAE as the displacements get smaller (MACE-MP-0, MACE-MPA-0, MACE-Alexandria, Mattersim, SevenNet). For ORB and OMat models, we see a drastic increase in phonon MAE as the displacements become smaller. This signifies that ORB and OMat models could possibly perform poorly in the low-force regime due to increased noise from the small displacements. This is consistent with recent results, where it was reported that OMat and ORB models had substantial errors in phonon predictions (for small displacements) when compared to models such as MatterSim, MACE, CHGNet, and SevenNet for 10,000 materials \cite{loew2024universalmachinelearninginteratomic}. This work hypothesized that the errors are due to ORB and OMat computing forces as a direct output of the neural network as opposed to calculating the forces by evaluating the derivative of the energy with respect to the atomic positions (through back-propagation) \cite{loew2024universalmachinelearninginteratomic}. MatterSim offers similar performance (and consistency across atomic displacements) in terms of predicting the phonon band structure when compared to accurate equivariant uMLFFs such as MACE and SevenNet. This is especially interesting since MatterSim-v1 is not an equivariant uMLFF. In addition, we see improved phonon results for the newer MACE-MPA-0 when compared to MACE-MP-0.

In addition to summing the computational timing for the entire workflow, we also examined the scaling behavior of each uMLFF (on CPU). These results for supercells of Cu (up to 10,000 atoms) are depicted in Fig. \ref{scalingandlb}a). Unsurprisingly, we see that OMat model scales much more drastically with system size as compared to the other uMLFF models. In addition, we found that the invariant models have more favorable scaling than the equivariant models such as MACE and SevenNet, with SevenNet having slightly better scaling than MACE. The scalability of ORB and MatterSim is a huge advantage due to their accuracy of various material property predictions. Similar scalability studies were conducted in Ref. \cite{neumann2024orbfastscalableneural} The quantities of surface energy and vacancy formation energy are difficult to predict with machine learning methods and provide a rigorous benchmarking test for uMLFF architectures. Out of the 104 materials, there exists 85 entries for surface DFT calculations and 48 entries for defect DFT calculations within JARVIS-DFT. We used these calculations to compare our uMLFF results. Fig. \ref{mlff-parity}a) and Fig. \ref{mlff-parity}b) depict the parity plots and corresponding MAE for surface energy and vacancy formation energy of our test sets. The very low error for surface energy (0.16 J/m$^2$) and vacancy formation energy (0.36 eV) highlight some of the successes of recent state-of-the-art models such as OMat24, ORB, MACE-MPA-0 and MatterSim. We also observe reasonably accurate results for MACE-MP-0 and SevenNet when predicting vacancy formation energy and surface energies. Recently, the proprietary model PFP from Preferred Networks Inc. was evaluated for our CHIPS-FF surface energy benchmark and achieved an MAE of 0.19 J/m$^2$ \cite{Iwase2025}, achieving similar accuracy to the OMat24 and ORB models. The low cost of ORB, high rate of convergence and high accuracy make it a viable tool to relax larger surfaces and defect supercells.

In addition to benchmarking the properties of bulk materials, we utilized the CHIPS-FF workflow for interface calculations. Fig. \ref{mlff-parity}c depicts the parity plot for work of adhesion (W$_{ad}$) for various material interfaces. The MAE for W$_{ad}$ is computed with respect to ``ground truth" data taken from Ref. \cite{intermat_interface_2024} (from experiment and theory). As seen in Fig. \ref{mlff-parity}c, predictions for W$_{ad}$ are extremely poor for each uMLFF type. The parity plot in Fig. \ref{mlff-parity}c is zoomed in to focus on a subset of calculations, but the MAE is computed for the entire test set of interfaces, which include significant outliers. This is not surprising due to the fact that none of these uMLFF are trained on interface data. We went on to analyze one of the best performing predictions for an Al(111)Al$_2$O$_3$(001) interface (see red box on Fig. \ref{mlff-parity}c). From here, we performed an in-plane (xy) grid search to find the optimal energy configuration of the interface. In addition, this in-plane scan can be a test of how smooth the potential energy surface is. On the right panel of Fig. \ref{mlff-parity}c we see the results of this interface scan for a number of uMLFFs. We observe a smoother potential energy surface for MACE, SevenNet, MatterSim and OMat24 models. Although the MAE is extremely high for all uMLFF models for W$_{ad}$, it is the lowest for ALIGNN-FF. This can be due in part to ALIGNN-FF being trained on vdW-corrected DFT calculations and having a substantial amount of exfoliable materials in the dataset.

We chose to benchmark the performance of uMLFF models for the generation of amorphous Si (a-Si) by performing melt/quench simulations. Fig. \ref{mlff-md} depicts a summary of these results, where the radial distribution function (RDF) is shown as a function of pair separation distance. We benchmarked these uMLFF RDF curves against the RDF curve obtained from computationally expensive AIMD simulations for a-Si. The inset of Fig. \ref{mlff-md} depicts the MAE of each RDF curve with respect to AIMD. We observe that architectures such as MACE, SevenNet, and OMat24 (eqV2), CHGNet,  ALIGNN-FF, MatterSim and ORB have excellent accuracy, while MatGL has a slightly higher MAE and a lower R$^2$ value. The final amorphous structures for AIMD, MatterSim, eqV2 and ORB (best performing uMLFF models) are given in the inset of Fig. \ref{mlff-md}. The exceptional performance of ORB and MatterSim (both invariant uMLFFs) for a-Si is significant due to the fact that it is much less costly than other similarly performing equivariant uMLFF models. The relatively high accuracy of ALIGNN-FF with regards to a-Si is surprising due to the fact that ALIGNN-FF had inferior performance for a number of properties and is also an invariant model. A more thorough assessment of how ALIGNN-FF and other uMLFFs can model amorphous materials will be the subject of future work, which will also utilize the CHIPS-FF infrastructure. These results bring into question whether or not equivariance is necessary to achieve accurate results for amorphous materials with uMLFFs, and more rigorous benchmarking will need to be conducted to bring clarity to this.

In addition to testing these uMLFF models on a smaller and more focused dataset, we also tested on a larger and more diverse dataset for force predictions. Unlike total energy, forces are more independent of exchange-correlation functional and the accuracy of force predictions on various datasets can give valuable insight into how a particular model performs. We chose to benchmark each uMLFF on the MLEARN dataset from Ref. \cite{mlearn}, which consists of face-centered cubic (Cu, Ni) and body-centered cubic (Li, Mo) metals and diamond group IV semiconductors (Si, Ge) which span a wide range of crystal structures and bonding environments ($\approx$ 200 - 300 data points for each element). These results are depicted in Table \ref{tab:comparison_metrics}. In addition, we benchmarked the accuracy of force predictions on very large datasets that were used to train uMLFFs. These datasets included ALIGNN\_FF\_DB (307,000 used to train ALIGNN-FF), MPF (188,000 used to train M3GNet), and MPTrj (1.58 million used to train CHGNet, MACE, SevenNet and used in the training of ORB and OMat models). These results are depicted in Table \ref{tab:comparison_metrics-more}. Unsurprisingly, superior accuracy is obtained across all datasets when the OMat and ORB models are used to predict forces. We also observe highly accurate forces for MatterSim and a substantial reduction in force error when going from MACE-MP-0 to MACE-MPA-0. The CHIPS-FF package has built in functions to perform these benchmarking calculations on each respective dataset, with a flexible framework to add additional datasets (such as OMat and Alexandria) and new uMLFFs as they develop over time.

\begin{table}[htbp]
  \centering
  \caption{A comparison of force errors (eV/\AA) and timings (s) for the mlearn DFT dataset for Mo, Li, Si, Ni, Cu, and Ge. The green color highlights the best performing model for each element.}
  \label{tab:comparison_metrics}
  \begin{adjustbox}{max width=\textwidth}
    \begin{tabular}{lccccccccccccc}
      \toprule
      & \multicolumn{2}{c}{Mo} & \multicolumn{2}{c}{Li} & \multicolumn{2}{c}{Si} & \multicolumn{2}{c}{Ni} & \multicolumn{2}{c}{Cu} & \multicolumn{2}{c}{Ge} \\
      \cmidrule(lr){2-3}\cmidrule(lr){4-5}\cmidrule(lr){6-7}\cmidrule(lr){8-9}\cmidrule(lr){10-11}\cmidrule(lr){12-13}
      \textbf{uMLFF Type} & Error (eV/\AA) & Time (s) & Error (eV/\AA) & Time (s) & Error (eV/\AA) & Time (s) & Error (eV/\AA)  & Time (s) & Error (eV/\AA) & Time (s) & Error (eV/\AA) & Time (s) \\
      \midrule
      alignn\_ff &
      0.869 & 1.2 x 10$^{2}$ &
      0.122 & 9.5 x 10$^{1}$ &
      0.324 & 9.5 x 10$^{1}$ &
      0.303 & 2.1 x 10$^{2}$ &
      0.216 & 1.9 x 10$^{2}$ &
      0.395 & 4.5 x 10$^{1}$ \\[6pt]

      chgnet &
      0.361 & 9.9 x 10$^{1}$ &
      0.045 & 1.4 x 10$^{2}$ &
      0.188 & 3.6 x 10$^{1}$ &
      0.059 & 3.1 x 10$^{2}$ &
      0.058 & 2.9 x 10$^{2}$ &
      0.207 & 3.3 x 10$^{1}$ \\[6pt]

      eqV2\_153M\_omat &
      0.096 & 3.0 x 10$^{3}$ &
      \cellcolor[HTML]{92D050}0.004 & 3.5 x 10$^{3}$ &
      \cellcolor[HTML]{92D050}0.012 & 6.6 x 10$^{3}$ &
      \cellcolor[HTML]{92D050}0.009 & 1.8 x 10$^{4}$ &
      0.006 & 9.2 x 10$^{3}$ &
      0.031 & 4.6 x 10$^{3}$ \\[6pt]

      eqV2\_31M\_omat &
      0.098 & 6.9 x 10$^{2}$ &
      \cellcolor[HTML]{92D050}0.004 & 7.9 x 10$^{2}$ &
      0.016 & 3.2 x 10$^{3}$ &
      0.010 & 3.4 x 10$^{3}$ &
      \cellcolor[HTML]{92D050}0.005 & 2.1 x 10$^{3}$ &
      0.033 & 1.1 x 10$^{3}$ \\[6pt]

      eqV2\_31M\_omat\_mp\_salex &
      0.056 & 6.9 x 10$^{2}$ &
      0.007 & 7.9 x 10$^{2}$ &
      0.021 & 2.2 x 10$^{3}$ &
      0.011 & 2.1 x 10$^{3}$ &
      0.006 & 3.9 x 10$^{3}$ &
      \cellcolor[HTML]{92D050}0.028 & 1.1 x 10$^{3}$ \\[6pt]

      eqV2\_86M\_omat &
      0.094 & 1.8 x 10$^{3}$ &
      \cellcolor[HTML]{92D050}0.004 & 2.0 x 10$^{3}$ &
      0.014 & 9.3 x 10$^{3}$ &
      \cellcolor[HTML]{92D050}0.009 & 9.9 x 10$^{2}$ &
      0.006 & 5.5 x 10$^{3}$ &
      0.032 & 2.7 x 10$^{3}$ \\[6pt]

      eqV2\_86M\_omat\_mp\_salex &
      \cellcolor[HTML]{92D050}0.048 & 1.8 x 10$^{3}$ &
      0.005 & 2.0 x 10$^{3}$ &
      0.016 & 4.9 x 10$^{3}$ &
      \cellcolor[HTML]{92D050}0.009 & 5.5 x 10$^{3}$ &
      \cellcolor[HTML]{92D050}0.005 & 9.7 x 10$^{3}$ &
      0.030 & 2.7 x 10$^{3}$ \\[6pt]

      mace &
      0.292 & 2.6 x 10$^{2}$ &
      0.035 & 2.8 x 10$^{2}$ &
      0.149 & 1.9 x 10$^{2}$ &
      0.055 & 1.8 x 10$^{3}$ &
      0.059 & 5.5 x 10$^{2}$ &
      0.203 & 1.9 x 10$^{2}$ \\[6pt]

      mace-alexandria &
      0.343 & 4.3 x 10$^{2}$ &
      0.025 & 4.7 x 10$^{2}$ &
      0.202 & 1.1 x 10$^{3}$ &
      0.099 & 2.2 x 10$^{3}$ &
      0.055 & 2.1 x 10$^{3}$ &
      0.161 & 9.4 x 10$^{2}$ \\[6pt]

      mace-mpa &
     0.130 &  2.1 x 10$^{2}$ &
     0.014 &  2.5 x 10$^{2}$ &
     0.083 &  3.5 x 10$^{2}$ &
     0.026 &  7.6 x 10$^{2}$ &
     0.013 &  7.4 x 10$^{2}$ &
     0.063 &  3.5 x 10$^{2}$ \\[6pt]

      matgl &
      0.333 & 1.0 x 10$^{1}$ &
      0.042 & 2.1 x 10$^{1}$ &
      0.226 & 4.1 x 10$^{1}$ &
      0.069 & 7.4 x 10$^{1}$ &
      0.124 & 6.5 x 10$^{1}$ &
      0.299 & 2.2 x 10$^{1}$ \\[6pt]

      matgl-direct &
      0.300 & 3.8 x 10$^{1}$ &
      0.031 & 3.9 x 10$^{1}$ &
      0.214 & 4.7 x 10$^{1}$ &
      0.114 & 1.6 x 10$^{2}$ &
      0.054 & 1.4 x 10$^{2}$ &
      0.260 & 4.8 x 10$^{1}$ \\[6pt]

      mattersim &
     0.121 & 1.2 x 10$^{2}$ &
     0.024  & 2.2 x 10$^{1}$ &
     0.067  & 1.5 x 10$^{2}$ &
     0.021  & 5.3 x 10$^{2}$ &
     0.012  & 4.6 x 10$^{2}$ &
     0.068  & 1.3 x 10$^{2}$ \\[6pt]

      orb-v2 &
      0.130 & 1.1 x 10$^{2}$ &
      0.018 & 1.2 x 10$^{2}$ &
      0.068 & 1.4 x 10$^{2}$ &
      0.026 & 2.8 x 10$^{2}$ &
      0.015 & 2.8 x 10$^{2}$ &
      0.060 & 1.4 x 10$^{2}$ \\[6pt]

      sevennet &
      0.365 & 1.9 x 10$^{2}$ &
      0.051 & 1.4 x 10$^{2}$ &
      0.129 & 1.4 x 10$^{2}$ &
      0.089 & 8.4 x 10$^{2}$ &
      0.055 & 3.7 x 10$^{2}$ &
      0.131 & 1.4 x 10$^{2}$ \\

      \bottomrule
    \end{tabular}
  \end{adjustbox}
\end{table}

\begin{table}[htbp]
  \centering
  \caption{A comparison of force errors (eV/\AA) and timings (s) for different datasets: ALIGNN\_FF\_DB, MPF, and MPTrj. The green color highlights the best performing model.}
  \label{tab:comparison_metrics-more}
  \begin{adjustbox}{max width=\textwidth}
    \begin{tabular}{lcccccc}
      \toprule
      & \multicolumn{2}{c}{ALIGNN\_FF\_DB (307k)} & \multicolumn{2}{c}{MPF (188k)} & \multicolumn{2}{c}{MPTrj (1.58M)} \\
      \cmidrule(lr){2-3}\cmidrule(lr){4-5}\cmidrule(lr){6-7}
      \textbf{MLFF Type} & Error (eV/\AA) & Time (s) & Error (eV/\AA) & Time (s) & Error (eV/\AA) & Time (s) \\
      \midrule
      alignn\_ff                 & 0.472 & 2.9 x 10$^{4}$ & 0.639 & 4.1 x 10$^{4}$ & 0.581 & 3.5 x 10$^{5}$ \\
      chgnet                     & 0.088 & 3.2 x 10$^{4}$ & 0.101 & 1.1 x 10$^{5}$ & 0.060 & 7.6 x 10$^{5}$ \\
      eqV2\_153M\_omat           & \cellcolor[HTML]{92D050}0.050 & 1.5 x 10$^{6}$ & - & - & - & - \\
      eqV2\_31M\_omat            & 0.051 & 3.0 x 10$^{5}$ & 0.046 & 3.5 x 10$^{5}$ & - & - \\
      eqV2\_31M\_omat\_mp\_salex & 0.055 & 3.9 x 10$^{5}$ & \cellcolor[HTML]{92D050}0.032 & 4.3 x 10$^{5}$ & - & - \\
      eqV2\_86M\_omat            & 0.052 & 9.2 x 10$^{5}$ & 0.043 & 9.3 x 10$^{5}$ & - & - \\
      eqV2\_86M\_omat\_mp\_salex & 0.056 & 9.2 x 10$^{5}$ & - & - & - & - \\
      mace                       & 0.104 & 5.0 x 10$^{4}$ & 0.125 & 1.8 x 10$^{5}$ & 0.059 & 1.1 x 10$^{6}$ \\
      mace-alexandria            & 0.221 & 1.3 x 10$^{5}$ & 0.395 & 2.7 x 10$^{5}$ & - & - \\
      mace-mpa            & 0.072 & 7.0 x 10$^{5}$ &  0.061 & 1.3 x 10$^{5}$ & 0.039 & 1.9 x 10$^{6}$  \\
      matgl                      & 0.091 & 1.1 x 10$^{4}$ & 0.066 & 3.4 x 10$^{4}$ & 0.100 & 1.4 x 10$^{5}$ \\
      matgl-direct               & 0.104 & 1.7 x 10$^{4}$ & 0.095 & 4.2 x 10$^{4}$ & 0.118 & 2.7 x 10$^{5}$ \\
      mattersim               & 0.071 & 8.7 x 10$^{4}$ & 0.047 & 7.1 x 10$^{4}$ & 0.059 & 1.7 x 10$^{6}$\\
      orb-v2                     & 0.062 & 5.0 x 10$^{4}$ & 0.049 & 8.4 x 10$^{4}$ & \cellcolor[HTML]{92D050}0.028 & 7.5 x 10$^{5}$ \\
      sevennet                   & 0.092 & 4.4 x 10$^{4}$ & 0.081 & 6.0 x 10$^{4}$ & 0.041 & 5.0 x 10$^{5}$ \\
      \bottomrule
    \end{tabular}
  \end{adjustbox}
\end{table}

We have introduced CHIPS-FF, an open-source benchmarking platform for MLFF architectures, which has the capability to test properties beyond the standard total energy, including forces, phonons, elastic properties, surface energy, vacancy formation energy and properties of interfaces. We benchmarked several recent state-of-the-art uMLFF architectures on a subset of 104 materials most relevant for the semiconductor industry, taking into account accuracy with respect to DFT, convergence and computational cost. As MLFFs continue to develop over time, we expect the CHIPS-FF benchmarking platform to be critical in terms of testing the quality of uMLFFs.

 \section{Data Availability Statement}
The CHIPS-FF package can be found at \url{https://github.com/usnistgov/chipsff}. All data specific to this work will be made available on Figshare upon publication. Related benchmarks for this work can be found at \url{https://pages.nist.gov/jarvis_leaderboard/Special/CHIPS_FF/}.

 \section{Conflicts of Interest}
The authors declare no competing interests.

 \section{Supporting Information}
Description of software used, detailed description of test set, results for ExpCellFilter, CHIPS-FF tutorial

\section{Acknowledgments}
All authors thank the National Institute of Standards and Technology for funding, computational, and data-management resources. This work was performed with funding from the CHIPS Metrology Program, part of CHIPS for America, National Institute of Standards and Technology, U.S. Department of Commerce. Certain commercial equipment, instruments, software, or materials are identified in this paper in order to specify the experimental procedure adequately. Such identifications are not intended to imply recommendation or endorsement by NIST, nor it is intended to imply that the materials or equipment identified are necessarily the best available for the purpose. The authors would like to acknowledge John Bonini (NIST), Orbital Materials, and Preferred Networks for fruitful discussions regarding the manuscript and software package.

\bibliography{main}

\end{document}


\maketitle

CHIPS-FF mainly relies on several Python-based packages: Atomic Simulation Environment (ASE) \cite{ase}, JARVIS-Tools \cite{10.1063/5.0159299,jarvis}, phonopy \cite{phono3py,phonopy-phono3py-JPCM}, elastic \cite{jochym_elastic,elastic1,elastic2} and Intermat \cite{D4DD00031E}. These packages work in conjunction to design and manipulate atomic structures, access materials databases and compute physically observable quantities. ASE is a general toolkit for setting up, running and analyzing atomistic calculations for a variety of electronic structure and molecular dynamics codes. In addition, it allows for the integration of machine learning force fields (MLFFs) as calculators within the atomistic workflows. JARVIS-Tools, which utilizes ASE for certain functionalities, is an infrastructure to perform high-throughput atomistic calculations, manipulate structures, and parse data and results from different electronic structure and molecular dynamics codes. The main feature of JARVIS-Tools is the seamless integration of large-scale materials databases, such as JARVIS-DFT, which can be used to generate initial structures, filter materials based on specific properties and compare data with density functional theory (DFT) benchmarks. In this work, we utilize the JARVIS-DFT datasets of bulk materials, surfaces, and defects. The phonopy package is used to determine phonon and lattice dynamics based on computed force constants, which can be obtained from the finite-displacement method. These force constants can be computed via DFT calculations or via MLFF. The phonon dispersion, phonon density of states and derived properties such as heat capacity and zero-point energy can be calculated from phonopy. The elastic package (which utilizes routines in ASE) is based on standard elasticity theory and the finite deformation approach to compute the elastic tensor of a crystal. The Intermat package is a toolkit for the generation of material interfaces that utilizes ASE and JARVIS-Tools to carry out high-throughput simulations. Interface configurations can be generated using the algorithm of Zur et al. \cite{10.1063/1.333084} and the x-y (in-plane) and z (out-of-plane) parameters of the interface can be optimized with DFT or MLFFs (passed through ASE).

\begin{figure}[h]
    \centering
    \includegraphics[trim={0. 0cm 0 0cm},clip,width=1.0\textwidth]{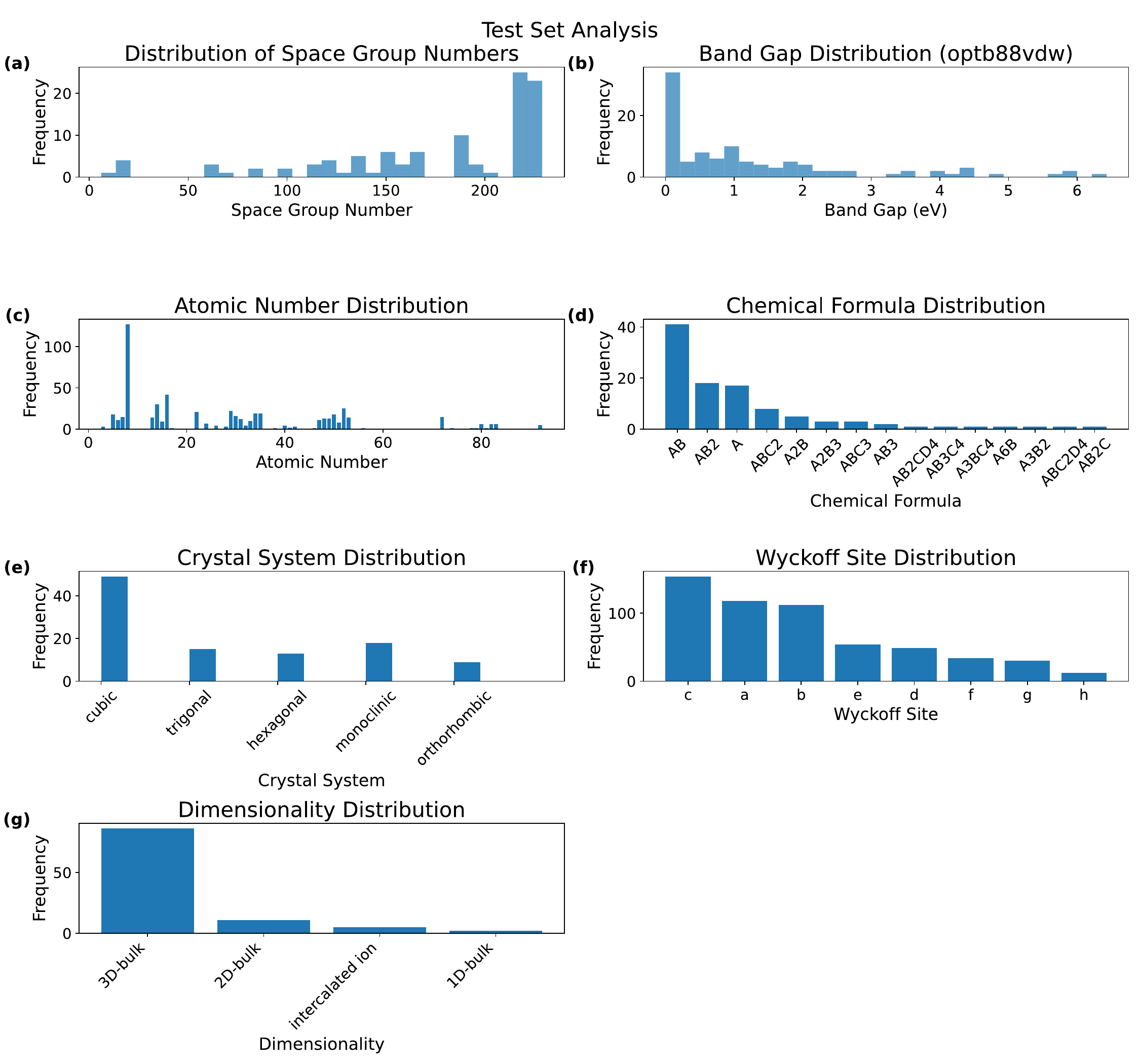}
    \caption{The data distribution of the 104 materials in the test set for a) space group number, b) band gap (vdW-DF-optB88), c) atomic number, d) chemical formula, e) crystal system, f) Wychkoff site, g) dimensionality. }
    \label{testset}
\end{figure}

\newpage

\begin{table}[h!]
\centering
\begin{adjustbox}{width=0.75\textwidth}
\begin{tabular}{l l l l l}
\hline
\textbf{JID} & \textbf{Formula} & \textbf{SPG Number} & \textbf{SPG Symbol} & \textbf{optb88vdw Band Gap (eV)} \\ \hline
JVASP-1002  & Si                & 227 & \textit{Fd-3m}     & 0.731 \\
JVASP-10036 & TiO$_2$         & 136 & \textit{P4$_2$/mnm}    & 1.769 \\
JVASP-10037 & SnO$_2$         & 136 & \textit{P4$_2$/mnm}    & 0.894 \\
JVASP-1008  & Sn                & 227 & \textit{Fd-3m}     & 0 \\
JVASP-1023  & Te                & 152 & \textit{P3$_{121}$}     & 0.165 \\
JVASP-1029  & Ti                & 191 & \textit{P6/mmm}    & 0 \\
JVASP-103127 & AlInSb$_2$     & 115 & \textit{P-4m2}     & 0.242 \\
JVASP-104   & TiO$_2$         & 141 & \textit{I4$_1$/amd}   & 2.023 \\
JVASP-104764 & Al$_3$GaN$_4$ & 6   & \textit{Pm}        & 3.505 \\
JVASP-105410 & SiGe             & 216 & \textit{F-43m}     & 0.694 \\
JVASP-10591 & ZnS               & 186 & \textit{P6$_3$mc}   & 2.102 \\
JVASP-106363 & InGaCu$_2$Se$_4$ & 82 & \textit{I-4} & 0.022 \\
JVASP-106686 & ZnHgTe$_2$     & 115 & \textit{P-4m2}     & 0 \\
JVASP-1067 & Bi$_2$Se$_3$   & 166 & \textit{R-3m}       & 0.319 \\
JVASP-107 & SiC                 & 186 & \textit{P6$_3$mc}  & 2.495 \\
JVASP-10703 & Cd$_3$As$_2$   & 224 & \textit{Pn-3m}     & 0 \\
JVASP-108770 & InGaSb$_2$     & 115 & \textit{P-4m2}     & 0 \\
JVASP-110 & BaTiO$_3$         & 99  & \textit{P4mm}      & 1.749 \\
JVASP-110231 & InGaN$_2$      & 156 & \textit{P3m1}      & 0.217 \\
JVASP-1103 & TePb               & 225 & \textit{Fm-3m}     & 1.119 \\
JVASP-1109 & SnS                & 62  & \textit{Pnma}      & 1.018 \\
JVASP-111005 & SnTe$_2$Pb     & 166 & \textit{R-3m}      & 0.488 \\
JVASP-1112 & PbS                & 225 & \textit{Fm-3m}     & 0.564 \\
JVASP-1115 & PbSe               & 225 & \textit{Fm-3m}     & 0.513 \\
JVASP-113 & ZrO$_2$           & 14  & \textit{P2$_1$/c} & 3.622 \\
JVASP-1174 & GaAs               & 216 & \textit{F-43m}     & 0.085 \\
JVASP-1177 & GaSb               & 216 & \textit{F-43m}     & 0 \\
JVASP-1180 & InN                & 186 & \textit{P6$_3$mc} & 0 \\
JVASP-1183 & InP                & 216 & \textit{F-43m}     & 0.331 \\
JVASP-1186 & InAs               & 216 & \textit{F-43m}     & 0 \\
JVASP-1189 & InSb               & 216 & \textit{F-43m}     & 0 \\
JVASP-1192 & CdSe               & 216 & \textit{F-43m}     & 0.455 \\
JVASP-1195 & ZnO                & 186 & \textit{P6$_3$mc} & 0.965 \\
JVASP-1198 & ZnTe               & 216 & \textit{F-43m}     & 1.057 \\
JVASP-1201 & CuCl               & 216 & \textit{F-43m}     & 0.729 \\
JVASP-1216 & Cu$_2$O          & 224 & \textit{Pn-3m}     & 0.644 \\
JVASP-1222 & UO$_2$           & 225 & \textit{Fm-3m}     & 0 \\
JVASP-1240 & LiNbO$_3$        & 161 & \textit{R3c}       & 3.339 \\
JVASP-131 & SnS$_2$           & 164 & \textit{P-3m1}     & 1.224 \\
JVASP-1312 & BP                 & 216 & \textit{F-43m}     & 1.515 \\
JVASP-1327 & AlP                & 216 & \textit{F-43m}     & 1.794 \\
JVASP-133719 & BAs              & 216 & \textit{F-43m}     & 1.296 \\
JVASP-1372 & AlAs               & 216 & \textit{F-43m}     & 1.681 \\
JVASP-1408 & AlSb               & 216 & \textit{F-43m}     & 1.322 \\
JVASP-14616 & Li                & 229 & \textit{Im-3m}     & 0 \\
JVASP-14968 & TiSi$_2$        & 70  & \textit{Fddd}      & 0 \\
JVASP-14970 & Si$_2$Mo        & 139 & \textit{I4/mmm}    & 0 \\
JVASP-149871 & GaAgS$_2$      & 122 & \textit{I-42d}     & 0.89 \\
JVASP-149906 & ZnCdTe$_2$     & 122 & \textit{I-42d}     & 0.68 \\
JVASP-149916 & SnTe             & 225 & \textit{Fm-3m}     & 0.479 \\
JVASP-18983 & TiO$_2$         & 61  & \textit{Pbca}      & 2.254 \\
\hline
\end{tabular}
\caption{A detailed description of the materials in the test set (part one), including JARVIS-ID, space group number and symbol and band gap (vdW-DF-optB88).}
\end{adjustbox}
\end{table}

\newpage
\begin{table}[h!]
\centering
\begin{adjustbox}{width=0.75\textwidth}
\begin{tabular}{l l l l l}
\hline
\textbf{JID} & \textbf{Formula} & \textbf{SPG Number} & \textbf{SPG Symbol} & \textbf{optb88vdw Band Gap (eV)} \\ \hline
JVASP-1915 & InSe               & 160 & \textit{R3m}       & 0.174 \\
JVASP-19780 & Si$_2$W         & 139 & \textit{I4/mmm}    & 0 \\
JVASP-20092 & CdO               & 225 & \textit{Fm-3m}     & 0 \\
JVASP-21211 & Se                & 152 & \textit{P3$_{121}$} & 0.898 \\
JVASP-22694 & NiO               & 225 & \textit{Fm-3m}     & 0 \\
JVASP-23 & CdTe                 & 216 & \textit{F-43m}     & 0.498 \\
JVASP-2376 & ZnSiP$_2$        & 122 & \textit{I-42d}     & 1.428 \\
JVASP-25 & Bi$_2$Te$_3$     & 166 & \textit{R-3m}      & 0.351 \\
JVASP-29539 & PbI$_2$         & 186 & \textit{P6$_3$mc} & 2.259 \\
JVASP-30 & GaN                  & 186 & \textit{P6$_3$mc} & 1.943 \\
JVASP-32 & Al$_2$O$_3$      & 167 & \textit{R-3c}      & 6.43 \\
JVASP-34249 & HfO$_2$         & 225 & \textit{Fm-3m}     & 4.037 \\
JVASP-34674 & SiO$_2$         & 20  & \textit{C222$_1$} & 5.673 \\
JVASP-3510 & BiI$_3$          & 148 & \textit{R-3}       & 2.365 \\
JVASP-36018 & GeC               & 216 & \textit{F-43m}     & 1.927 \\
JVASP-36408 & SnC               & 216 & \textit{F-43m}     & 0.748 \\
JVASP-36873 & BSb               & 216 & \textit{F-43m}     & 0.945 \\
JVASP-39 & AlN                  & 186 & \textit{P6$_3$mc} & 4.474 \\
JVASP-41 & SiO$_2$            & 154 & \textit{P3$_{221}$} & 5.986 \\
JVASP-4282 & CrBr$_3$         & 148 & \textit{R-3}       & 1.299 \\
JVASP-43367 & HfO$_2$         & 61  & \textit{Pbca}      & 4.033 \\
JVASP-5224 & HgI$_2$          & 137 & \textit{P4$_2$/nmc} & 0.969 \\
JVASP-54 & MoS$_2$            & 194 & \textit{P6$_3$/mmc} & 0.922 \\
JVASP-58349 & SiO$_2$         & 152 & \textit{P3$_{121}$} & 5.985 \\
JVASP-62940 & BN                & 194 & \textit{P6$_3$/mmc} & 4.46 \\
JVASP-7836 & BN                 & 216 & \textit{F-43m}     & 4.813 \\
JVASP-79522 & CuO               & 131 & \textit{P4$_2$/mmc} & 0 \\
JVASP-8003 & CdS                & 216 & \textit{F-43m}     & 0.993 \\
JVASP-802 & Hf                  & 194 & \textit{P6$_3$/mmc} & 0 \\
JVASP-8082 & SrTiO$_3$        & 221 & \textit{Pm-3m}     & 1.814 \\
JVASP-8118 & SiC                & 186 & \textit{P6$_3$mc} & 2.618 \\
JVASP-813 & Ag                  & 225 & \textit{Fm-3m}     & 0 \\
JVASP-8158 & SiC                & 216 & \textit{F-43m}     & 1.62 \\
JVASP-816 & Al                  & 225 & \textit{Fm-3m}     & 0 \\
JVASP-8184 & GaP                & 186 & \textit{P6$_3$mc} & 1.263 \\
JVASP-825 & Au                  & 225 & \textit{Fm-3m}     & 0 \\
JVASP-85416 & Ag$_2$S         & 14  & \textit{P2$_1$/c} & 1.11 \\
JVASP-85478 & Cu$_2$S         & 96  & \textit{P4$_{32}$/12} & 0.59 \\
JVASP-8554 & InCuSe$_2$       & 122 & \textit{I-42d}     & 0.006 \\
JVASP-8559 & TlBr               & 221 & \textit{Pm-3m}     & 2.023 \\
JVASP-861 & Cr                  & 229 & \textit{Im-3m}     & 0 \\
JVASP-867 & Cu                  & 225 & \textit{Fm-3m}     & 0 \\
JVASP-890 & Ge                  & 227 & \textit{Fd-3m}     & 0 \\
JVASP-90668 & ZnCu$_2$SnS$_4$ & 82 & \textit{I-4}    & 0.115 \\
JVASP-91 & C                    & 227 & \textit{Fd-3m}     & 4.457 \\
JVASP-9117 & FeS$_2$          & 205 & \textit{Pa-3}      & 0.433 \\
JVASP-9147 & HfO$_2$          & 14  & \textit{P2$_1$/c} & 4.123 \\
JVASP-9166 & B$_6$As          & 166 & \textit{R-3m}      & 2.739 \\
JVASP-943 & Ni                  & 225 & \textit{Fm-3m}     & 0 \\
JVASP-96 & ZnSe                 & 216 & \textit{F-43m}     & 1.224 \\
JVASP-963 & Pd                  & 225 & \textit{Fm-3m}     & 0 \\
JVASP-972 & Pt                  & 225 & \textit{Fm-3m}     & 0 \\
JVASP-99732 & CdHg$_3$Te$_4$ & 215 & \textit{P-43m}  & 0 \\ 
\hline
\end{tabular}
\caption{A detailed description of the materials in the test set (part two), including JARVIS-ID, space group number and symbol and band gap (vdW-DF-optB88).}
\end{adjustbox}
\end{table}

\begin{table}[htbp]
  \centering
  \caption{The percentage of unconverged structural relaxations for bulk, surface, and vacancy calculations for each uMLFF (relaxed using the \texttt{ExpCellFilter}). The green highlights which models have the highest convergence rate. }
  \label{tab:bulk_surface_vacancy}
  \begin{adjustbox}{max width=\textwidth}
    \begin{tabular}{lccc}
      \toprule
      \textbf{uMLFF Type} & \textbf{Bulk} & \textbf{Surface} & \textbf{Vacancy} \\
      \midrule

    alignn\_ff                 & 7 & 38 & 38 \\
      chgnet                     & \cellcolor[HTML]{92D050}0  & 1 & \cellcolor[HTML]{92D050}0 \\
      eqV2\_153M\_omat           & \cellcolor[HTML]{92D050}0  & 15 & 2 \\
      eqV2\_31M\_omat            & \cellcolor[HTML]{92D050}0  & 17 & \cellcolor[HTML]{92D050}0 \\
      eqV2\_31M\_omat\_mp\_salex & \cellcolor[HTML]{92D050}0  & 6 & \cellcolor[HTML]{92D050}0 \\
      eqV2\_86M\_omat            & 2  & 18 & 2 \\
      eqV2\_86M\_omat\_mp\_salex & \cellcolor[HTML]{92D050}0  & 6 & \cellcolor[HTML]{92D050}0 \\
      mace                       & \cellcolor[HTML]{92D050}0  & 7 & 4 \\
      mace-alexandria            & 3  & 16 & 2 \\
      mace-mpa                       & \cellcolor[HTML]{92D050}0  & 7 & \cellcolor[HTML]{92D050}0 \\
      matgl                      & \cellcolor[HTML]{92D050}0  & 5 & 10 \\
      matgl-direct               & 4  & 7 & 13 \\
      mattersim               & \cellcolor[HTML]{92D050}0  & 5 & 2 \\
      orb-d3-v2                     & \cellcolor[HTML]{92D050}0  & 1 & \cellcolor[HTML]{92D050}0 \\
      orb-v2                     & 1  & \cellcolor[HTML]{92D050}0 & \cellcolor[HTML]{92D050}0 \\
      sevennet                   & \cellcolor[HTML]{92D050}0  & 2 & \cellcolor[HTML]{92D050}0 \\
      \bottomrule
    \end{tabular}
  \end{adjustbox}
\end{table}

\begin{table}[htbp]
  \centering
  \caption{The mean absolute error (MAE) for lattice constants a, b, c, and volume calculated with each uMLFF type (relaxed using the \texttt{ExpCellFilter}), compared to JARVIS-DFT data. The green highlights the best performing models.}
  \label{tab:error_metrics}
  \begin{adjustbox}{max width=\textwidth}
    \begin{tabular}{lcccc}
      \toprule
      \textbf{uMLFF Type} & \textbf{err\_a (\AA)} & \textbf{err\_b (\AA)} & \textbf{err\_c (\AA)} & \textbf{err\_vol (\AA$^{3}$)} \\
      \midrule
      
      alignn\_ff                & 0.087  & 0.102 & 0.137 & 9.95 \\
      chgnet                     & 0.046  & 0.049 & 0.109 & 3.58 \\
      eqV2\_153M\_omat           & 0.036  & 0.040 & 0.111 & 2.90 \\
      eqV2\_31M\_omat            & 0.029  & 0.032 & 0.099 & 3.05 \\
      eqV2\_31M\_omat\_mp\_salex & 0.028  & 0.030 & 0.096 & 3.17 \\
      eqV2\_86M\_omat            & 0.033  & 0.040 & 0.094 & 3.41 \\
      eqV2\_86M\_omat\_mp\_salex & 0.027  & 0.030 & 0.100 & 3.14 \\
      mace                       & 0.035  & 0.038 & 0.084 & 3.00 \\
      mace-alexandria            & 0.081  & 0.086 & 0.206 & 6.05 \\
      mace-mpa                   & 0.044  & 0.046 & 0.098 & 3.57 \\
      matgl                      & 0.052  & 0.057 & 0.128 & 3.50 \\
      matgl-direct               & 0.044  & 0.046 & 0.106 & 2.88 \\
      mattersim                  & 0.031  & 0.033 & 0.110 & 3.18 \\
      orb-d3-v2                  & 0.031  & 0.030 & \cellcolor[HTML]{92D050}0.048 & \cellcolor[HTML]{92D050}1.90 \\
      orb-v2                     & \cellcolor[HTML]{92D050}0.023  & \cellcolor[HTML]{92D050}0.025 & 0.159 & 3.38 \\
      sevennet                   & 0.035  & 0.038 & 0.100 & 3.43 \\
      
      \bottomrule
    \end{tabular}
  \end{adjustbox}
\end{table}

\newpage

\section{Tutorial}

\subsection{Installation and Setup}
To install CHIPS-FF, first clone the repository and set up a conda environment. 

\begin{codeblock}[title={Cloning and Setting Up the Environment}]
git clone https://github.com/usnistgov/chipsff
conda env create -f environment.yml -n chipsff
conda activate chipsff
cd chipsff
pip install -e .
\end{codeblock}

\subsection{Input Configuration}
CHIPS-FF utilizes a JSON input file that specifies various parameters such as the material identifier, calculator type, and simulation settings. Below is an example input file (\texttt{input.json}) for a single material:

\begin{codeblock}[title={Example input.json}]
{
  "jid": "JVASP-1002",
  "calculator_type": "chgnet",
  "chemical_potentials_file": "chemical_potentials.json",
  "properties_to_calculate": [
    "relax_structure",
    "calculate_ev_curve",
    "calculate_formation_energy",
    "calculate_elastic_tensor",
    "run_phonon_analysis",
    "analyze_surfaces",
    "analyze_defects",
    "run_phonon3_analysis",
    "general_melter",
    "calculate_rdf"
  ],
  "bulk_relaxation_settings": {
    "filter_type": "FrechetCellFilter",
    "relaxation_settings": {
      "fmax": 0.05,
      "steps": 200,
      "constant_volume": false
    }
  },
  "phonon_settings": {
    "dim": [2, 2, 2],
    "distance": 0.01
  },
  "use_conventional_cell": false
}
\end{codeblock}

Here is an example input file for material interfaces (e.g., \texttt{interface\_input.json}):

\begin{codeblock}[title={Example interface\_input.json}]
{
  "film_id": ["JVASP-1002"],
  "substrate_id": ["JVASP-816"],
  "calculator_type": "alignn_ff",
  "chemical_potentials_file": "chemical_potentials.json",
  "film_index": "1_1_0",
  "substrate_index": "1_1_0",
  "properties_to_calculate": [
    "analyze_interfaces"
  ]
}
\end{codeblock}

\subsection{Running the Simulations}
The main \texttt{run\_chipsff.py} script allows for a command-line interface for executing the simulations. For a single material analysis:

\begin{codeblock}[title={Running Single Material Analysis}]
python run_chipsff.py --input_file input.json
\end{codeblock}

For interface analysis:

\begin{codeblock}[title={Running Interface Analysis}]
python run_chipsff.py --input_file interface_input.json
\end{codeblock}

\subsection{Overview of Key Methods}
The CHIPS-FF framework includes several key functions:
\begin{itemize}
  \item \texttt{relax\_structure()}: Optimizes atomic structures.
  \item \texttt{calculate\_formation\_energy()}: Computes formation energies using relaxed structures and chemical potential data.
  \item \texttt{calculate\_elastic\_tensor()}: Evaluates elastic properties.
  \item \texttt{calculate\_ev\_curve()}: Fits energy-volume curves to determine equilibrium parameters.
  \item \texttt{run\_phonon\_analysis()}: Performs phonon band structure and thermal property calculations.
  \item \texttt{analyze\_defects()} and \texttt{analyze\_surfaces()}: Analyze defect and surface energies.
  \item \texttt{analyze\_defects\_from\_db()} Analyze defect energies with initial structures directly from the JARVIS-DFT vacancy database.
  \item \texttt{run\_phonon3\_analysis()}: Calculates third order force constants for thermal conductivity.
  \item \texttt{general\_melter()}: Executes MD simulations for melting and quenching processes.
  \item \texttt{analyze\_interfaces()}: Conducts interface analysis between film and substrate materials.
\end{itemize}

\bibliography{si}